%% file: main.tex
\documentclass[sigconf]{acmart}
\usepackage{booktabs,multirow}
\usepackage{enumitem}
\usepackage[normalem]{ulem}
\usepackage{kotex}
\usepackage{color}
\usepackage{rotating}
\usepackage{graphicx}
\usepackage{sidecap}
\usepackage[linesnumbered,algoruled,boxed,lined]{algorithm2e}

\newcommand{\ie}{{\it i.e.}}
\newcommand{\eg}{{\it e.g.}}
\newcommand{\com}{\textcolor{red}}

\newcommand{\fcom}{\textcolor{blue}}
\newcommand{\topN}{top-\emph{N}}
\newcommand{\argmin}{\operatornamewithlimits{argmin}}

\useunder{\uline}{\ul}{}

\copyrightyear{2022}
\acmYear{2022}
\setcopyright{acmlicensed}
\acmConference[WSDM '22] {Proceedings of the Fifteenth ACM International Conference on Web Search and Data Mining}{February 21--25, 2022}{Tempe, AZ, USA.}
\acmBooktitle{Proceedings of the Fifteenth ACM International Conference on Web Search and Data Mining (WSDM '22), February 21--25, 2022, Tempe, AZ, USA}
\acmPrice{15.00}
\acmISBN{978-1-4503-9132-0/22/02}
\acmDOI{10.1145/3488560.3498464}

\begin{document}
\fancyhead{}
\title{S-Walk: Accurate and Scalable Session-based Recommendation with Random Walks}

\author{Minjin Choi}
\affiliation{
  \institution{Sungkyunkwan University}
  \country{Republic of Korea}}
\email{zxcvxd@skku.edu}

\author{Jinhong Kim}
\affiliation{
  \institution{Naver Corp.}
  \country{Republic of Korea}}
\email{jinhong.kim93@navercorp.com}

\author{Joonseok Lee}
\affiliation{
  \institution{Google Research,\\Seoul National University}
  \country{United States, Republic of Korea}}
\email{joonseok2010@gmail.com}

\author{Hyunjung Shim}
\affiliation{
  \institution{Yonsei University}
  \country{Republic of Korea}}
\email{kateshim@yonsei.ac.kr}

\author{Jongwuk Lee}\authornote{Corresponding author}
\affiliation{
  \institution{Sungkyunkwan University}
  \country{Republic of Korea}}
\email{jongwuklee@skku.edu}

\input{sec-abstract}
\maketitle

\input{sec-introduction}
\input{sec-preliminaries}

\input{sec-model}
\input{sec-experiments}
\input{sec-results}

\input{sec-relatedwork}
\input{sec-conclusion}

\bibliographystyle{ACM-Reference-Format}
\bibliography{references}

\input{sec-appendix}

\end{document}

%% file: sec-abstract.tex
\begin{abstract}
Session-based recommendation (SR) predicts the next items from a sequence of previous items consumed by an anonymous user. Most existing SR models focus only on modeling \emph{intra-session} characteristics but pay less attention to \emph{inter-session} relationships of items, which has the potential to improve accuracy. Another critical aspect of recommender systems is computational efficiency and scalability, considering practical feasibility in commercial applications. To account for both accuracy and scalability, we propose a novel session-based recommendation with a random walk, namely \emph{S-Walk}. Precisely, S-Walk effectively captures intra- and inter-session correlations by handling high-order relationships among items using random walks with restart (RWR). By adopting linear models with closed-form solutions for transition and teleportation matrices that constitute RWR, S-Walk is highly efficient and scalable. Extensive experiments demonstrate that S-Walk achieves comparable or state-of-the-art performance in various metrics on four benchmark datasets. Moreover, the model learned by S-Walk can be highly compressed without sacrificing accuracy, conducting two or more orders of magnitude faster inference than existing DNN-based models, making it suitable for large-scale commercial systems.
\end{abstract}

\ccsdesc[500]{Information systems~Recommender systems}
\ccsdesc[500]{Information systems~Collaborative filtering}
\ccsdesc[300]{Information systems~Expert systems}

\keywords{Collaborative filtering; Session-based recommendation; Random walks; Closed-form solution}

%% file: sec-introduction.tex
\section{Introduction}

Modern recommender systems (RS) are indispensable for addressing the enormous information overload in various real-world applications, such as e-commerce platforms and online multimedia platforms, \eg, Amazon and Alibaba, YouTube, Netflix, and Spotify. Classical recommender systems~\cite{GoldbergNOT92, HerlockerKBR99, HuKV08, RicciRS15} usually assume that user accounts and users' long-term interactions are available. However, this assumption rarely holds. The user may not login, or multiple users share the user account, \eg, family members. The user may also exhibit different behaviors depending on the context. Thus, it is necessary to provide personalized recommendations without explicit user information.

Recently, session-based recommendation (SR)~\cite{BonninJ14,JannachLL17,WangCW19,FangGZS19,QuadranaCJ18} has gained considerable attention for predicting the next items from the sequential behavior consumed by an anonymous user. Unlike conventional RS, SR relies only on the users' actions in an ongoing session. This setting of SR is well-suited for real-world scenarios but inherently results in a severe data sparsity problem. To address this issue, it is essential to understand the unique characteristics of sessions, that is, \emph{intra-session} properties. First, the items within a session are coherent with the user's hidden intent, \eg, a list of products in the same category, referred to as \emph{item consistency} (or \emph{long-term dependency}). Second, some items are strictly consumed in chronological order, namely \emph{sequential dependency} (or \emph{short-term dependency}), \eg, consecutive episodes of a TV series. Lastly, the user can repeatedly consume the same items, called \emph{repeated item consumption}, \eg, user's favorite tracks.

\input{Figures/Fig1_session_relationship}

Most SR models utilize deep neural networks (DNNs) to learn intra-session relationships. Recurrent neural networks (RNNs)~\cite{HidasiKBT15, HidasiQKT16, HidasiK18} and attention mechanisms~\cite{LiRCRLM17, LiuZMZ18} have been used to model the sequential dependency of items. Recently, graph neural networks (GNNs)~\cite{WuT0WXT19,XuZLSXZFZ19,GuptaGMVS19,WangRMCMR19,PanCLR20a,Abu2020,PanCCCR20} have been used to effectively represent both item consistency and sequential dependency. Unfortunately, they suffer from performance degradation when dealing with complex and long sessions, where it is difficult to understand user intent. For this, \emph{inter-session} relationships among items are valuable clues. Figure~\ref{fig:session_relationship} describes intra-session and inter-session relationships; some items do not occur within a single session but share their neighboring items for multiple sessions, implying potential item correlations. Several studies~\cite{WangRMCMR19,LuoZLZWXFS20,WangWCLMQ20} have attempted to consider inter-session relationships using neighboring sessions or global item graphs. However, they incur substantial computational costs and are infeasible for large-scale commercial systems.

High-capacity DNN models have achieved state-of-the-art performances but usually require heavy computational overhead for runtime speed and memory consumption. Although competing for the best performance is a reasonable mission for most research problems, computational efficiency and scalability are also critical factors for dealing with practical restrictions in commercial recommender systems. \cite{LudewigJ18,GargGMVS19} suggested neighborhood-based models for session-based recommendations. Owing to their simplicity, they are highly scalable. Moreover, \cite{LudewigMLJ19a, LudewigMLJ19b} reported that the neighborhood-based models achieved comparable performances to DNN-based models on several benchmark datasets.

Our primary objective is to design an SR model that accounts for both accuracy and scalability. To this end, we propose a novel session-based recommendation with a random walk, namely \emph{S-Walk}, (i) exploiting intra- and inter-session relationships among items to improve accuracy, and (ii) supporting cost-effective, real-time performance at scale.

(i) Whereas the basic SR learns hidden patterns between items only within a session, S-walk additionally introduces \emph{global item graphs}, modeling item-item relationships across all sessions. By applying \emph{random walks with restart (RWR)} on this graph, a random surfer can jump from one item to another adjacent item by traversing the item graph or restart from an arbitrary item in the current session. Therefore, S-Walk can capture high-order correlations among items using multi-hop connections on an item graph. S-Walk can exploit the local patterns within a session and global patterns involving the same items from other sessions.

(ii) Recently, linear item-item models~\cite{Steck19a,Steck19b,Steck19c,JeunenBG20} have shown competitive performance in conventional RS. Motivated by their success, we devise linear item models to build two probability matrices, \emph{item transition} and \emph{item teleportation} matrices, to formulate the stochastic process of RWR. The item transition matrix captures the sequential dependency of items, generalizing a Markov chain model on items. The item teleportation matrix reflects the restart probability, allowing various items to participate in the model training depending on the ongoing session. Instead of having arbitrary items for restart, we utilize the co-occurrence relationship among items. The items that co-occurred with the current session items are used for restart. Notably, training our linear models is highly efficient and scalable because they have closed-form solutions whose computational complexity is determined by the number of items, independent of the number of sessions or user actions.

To summarize, the key advantages of S-Walk are as follows: (i) It can effectively capture inter- and intra-session relationships via RWR. (ii) Without complicated model tuning, it is highly efficient and scalable owing to the closed-form solution of linear models. (iii) It achieves competitive or state-of-the-art performance in various metrics (\ie, HR, MRR, recall, and MAP) on four benchmark datasets (\ie, YouChoose, Diginetica, RetailRocket, and NowPlaying). (iv) The model learned by S-Walk can be highly compressed without sacrificing accuracy, supporting fast inference time.

%% file: Figures/Fig1_session_relationship.tex
\begin{figure}
  \includegraphics[width=0.39\textwidth]{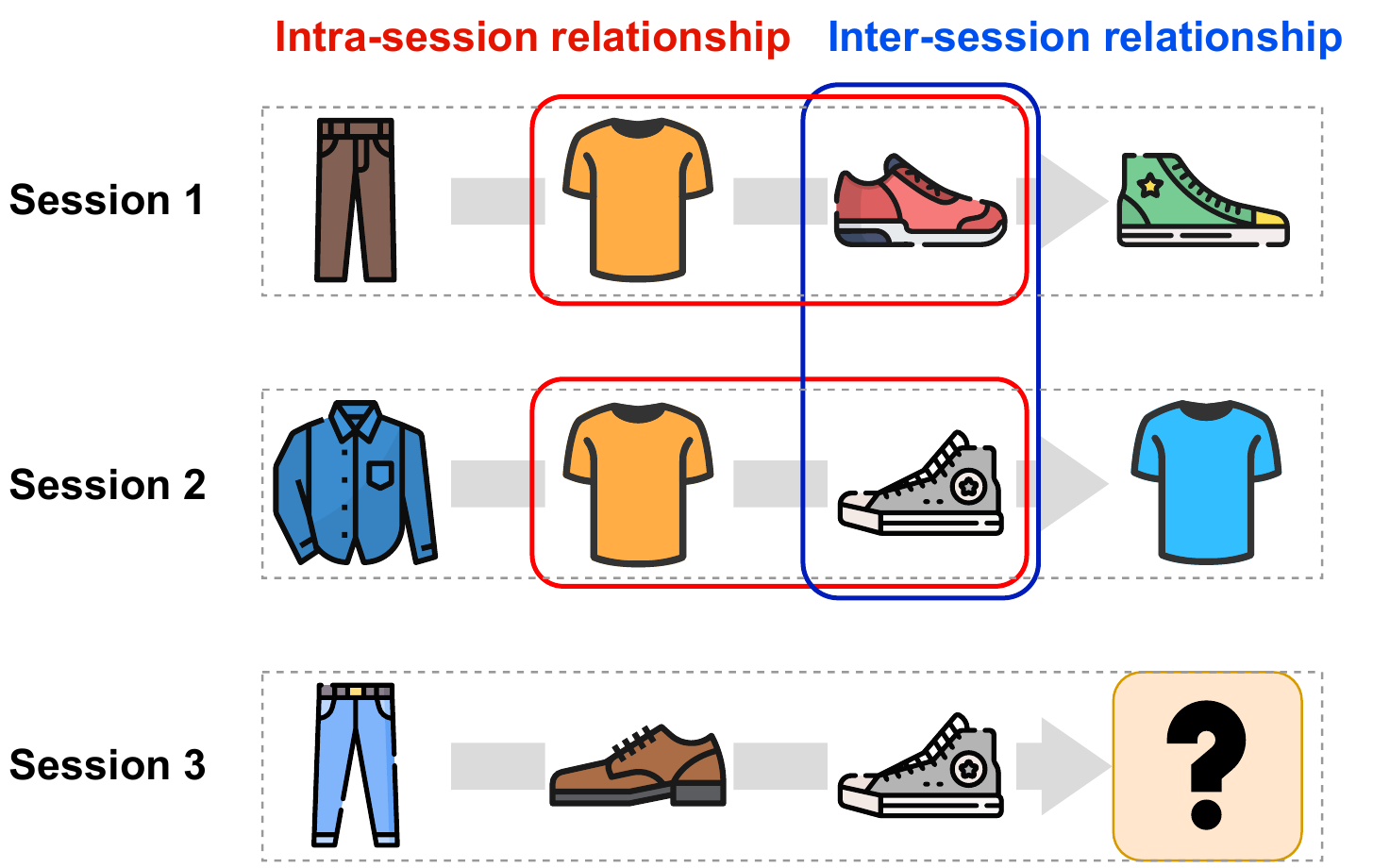}
  \caption{Illustration of intra- and inter-session relationships. The red shoes and the gray shoes (in blue box) do not appear within a single session, but they are correlated because both are related to the orange shirt (in red boxes).}
  \label{fig:session_relationship}
  \vskip -0.2in
\end{figure}

%% file: sec-preliminaries.tex
\section{Preliminaries}\label{sec:preliminaries}

\textbf{Notations.} Given a set of sessions $\mathcal{S} = \{s^{(1)}, \dots, s^{(m)}\}$ over a set of items $\mathcal{I} = \{i_1, \ldots, i_n\}$, an arbitrary session $s \in \mathcal{S}$ is represented by a sequence of items $s = (s_1, s_2, \dots, s_{|s|})$ with a length $|s|$. Here, $s_j \in \mathcal{I}$ is the $j$-th consumed item, \eg, clicked, watched, or purchased. For simplicity, $s \in \mathcal{S}$ is represented by a binary vector $\mathbf{x} \in \{0, 1\}^{n}$, where $\mathbf{x}_k = 1$ if $i_k$ is consumed, and $\mathbf{x}_k = 0$ otherwise. By stacking $m$ sessions, let $\mathbf{X} \in \mathbb{R}^{m \times n}$ denote a session-item interaction matrix, where $m$ is the number of sessions. As a straightforward variant, the binary value in $\mathbf{x}_k$ can be converted to real values to quantify the importance of items within a session.

\vspace{1mm}
\noindent
\textbf{Problem statement.} Given a sequence of items previously consumed by an anonymous user in a session, session-based recommendations predict the next items that the user is most likely to consume. Formally, a session-based recommender model takes a session $s = (s_1, \dots, s_j)$ as input and returns a ranked list of top-$N$ candidate items as the recommended next items $(s_{j+1}, \ldots s_{|s|})$. Note that this is more generalized (and challenging) than predicting the next single item $s_{j+1}$, which has been used in existing work. (See Section~\ref{sec:setup} for the generalized evaluation metrics for this setting.)

\vspace{1mm}
\noindent
\textbf{Random walk models.} The key concept behind random walk models is to reflect the direct and transitive relations among items. Conventional recommender models using random walks~\cite{YildirimK08,JamaliE09,CooperLRS14,ChristoffelPNB15,EksombatchaiJLL18,NikolakopoulosK19,NikolakopoulosK20} are based on a user-item bipartite graph $\mathcal{G}$ = ($\mathcal{U} \cup \mathcal{I}$, $\mathcal{E}$), where $\mathcal{U}$ and $\mathcal{I}$ are a set of users and items. Each edge $e \in \mathcal{E}$ represents the relationship between a user and an item. Thus, the core part of the random walk model is to determine a transition probability matrix to compute \emph{proximity} scores for items.

In general, there are two possible solutions for computing the proximity scores of items. First, we can utilize the $k$-step landing probability distribution of a random walker. Starting from a source user $u$, the proximity scores for all items can be computed as $\mathbf{u} \mathbf{R}_{(k)}$, where $\mathbf{u}$ is the user vector, and $\mathbf{R}_{(k)}$ is the $k$-step transition probability matrix. Although it is simple yet effective, existing studies~\cite{CooperLRS14, ChristoffelPNB15} are vulnerable to popularity bias; popular items tend to have high proximity scores as $k$ increases (\eg, $k \ge 3$), thereby degrading the recommendation quality.

As an alternative, we can compute the stationary distribution of \emph{random walks with restart (RWR)}, which is well-known as personalized PageRank~\cite{AmyNCarlD06pagerank}. It is effective for capturing high-order relationships among vertices. Because RWR leverages \emph{restart} in addition to \emph{sequential transition}, it can alleviate the problem of popularity bias, concentrating on the central node in the $k$-step. For this reason, we adopt RWR for session-based recommendations. (See Section~\ref{sec:results} for empirical comparisons between the two methods, $k$-step and RWR.)

Adopting the random walk in session-based recommendations has the following advantages: (i) The random walk model utilizes high-order item correlations across sessions. Because sessions are usually sparse by nature, it is useful for alleviating the data sparsity problem by capturing profound relationships among items. (ii) Compared to GNN-based SR models~\cite{WuT0WXT19,XuZLSXZFZ19,GuptaGMVS19,PanCLR20a,PanCCCR20}, it is efficient without requiring complicated hyper-parameter tuning.

\vspace{1mm}
\noindent
\textbf{Linear item-item models.} Given the session-item matrix $\mathbf{X}$, the goal of linear models~\cite{NingK11, Steck19b} is to estimate the item-item similarity matrix $\mathbf{B} \in \mathbb{R}^{n \times n}$. As a pioneering work, SLIM~\cite{NingK11} formulated a linear model subject to the constraint that all entries in $\mathbf{B}$ are non-negative and zero diagonal.
\begin{equation}
  \label{eq:slim_loss}
  \begin{aligned}
    \argmin_{\mathbf{B}} & \|\mathbf{X} - \mathbf{X} \cdot \mathbf{B}\|_F^2 + \lambda_1  \|\mathbf{B}\|_1 + \lambda_2  \|\mathbf{B}\|_F^2 \\
    & \text{s.t.} \ \ \texttt{diag}(\mathbf{B}) = 0, \ \ \mathbf{B} \ge 0,
  \end{aligned}
\end{equation}
where $\|\cdot\|_1$ and $\|\cdot\|_F$ are the entry-wise $\ell_1$-norm and the matrix Frobenius norm, respectively, $\lambda_1$ and $\lambda_2$ are the regularization coefficients, and $\texttt{diag}(\mathbf{B}) \in \mathbb{R}^{n}$ denotes the vector with the diagonal elements of $\mathbf{B}$. Although SLIM~\cite{NingK11} shows competitive accuracy, it suffers from high computational training cost.

Recently, EASE$^{R}$~\cite{Steck19b} and its variants~\cite{Steck19a,Steck19c} only consider the zero-diagonal constraint by removing the non-negativity of $\mathbf{B}$ and $\ell_1$-norm constraints from Eq.~\eqref{eq:slim_loss}:
\begin{equation}
  \label{eq:ease_loss}
  \argmin_{\mathbf{B}} \|\mathbf{X} - \mathbf{X} \cdot \mathbf{B}\|_F^2 + \lambda \cdot \|\mathbf{B}\|_F^2 \ \ \ \text{s.t.} \ \ \texttt{diag}(\mathbf{B}) = 0.
\end{equation}
Owing to this simpler formulation, EASE$^{R}$ is solved by the closed-form equation via Lagrange multipliers: \begin{equation}
  \label{eq:ease_solution}
  \hat{\mathbf{B}} = \mathbf{I} - \hat{\mathbf{P}} \cdot  \texttt{diagMat}(\textbf{1} \oslash \texttt{diag}(\hat{\mathbf{P}})),
\end{equation}
where $\hat{\mathbf{P}} = (\mathbf{X}^{\top} \mathbf{X} + \lambda \mathbf{I})^{-1}$. Here, $\textbf{1} \in \mathbb{R}^n$ is the vector of ones, and $\oslash$ denotes the element-wise division. Let $\texttt{diagMat}(\mathbf{x})$ denote the diagonal matrix expanded from the vector $\mathbf{x}$. (See~\cite{Steck19b} for the detailed derivation of the closed-form solution.)

Although inverting the regularized Gram matrix $\mathbf{X}^{\top} \mathbf{X}$ is the computational bottleneck for large-scale datasets (\ie, the time complexity is $\mathcal{O}(|\mathcal{I}|^{2.376})$ with the Coppersmith-Winograd algorithm), the closed-form solution is advantageous in terms of efficiency. The training complexity of EASE$^{R}$~\cite{Steck19b} is proportional to the number of items, which is usually much smaller than the number of sessions ($n \ll m$). Besides, the linear model is beneficial to accelerate the inference because computing \topN\ recommended items is simply done by single matrix multiplication. Most recently, SLIST~\cite{ChoiKLSL21SLIST} reported competitive accuracy with linear models for the session-based recommendation. Although SLIST~\cite{ChoiKLSL21SLIST} tackled various characteristics of session data, it did not consider inter-session relationships. In contrast, we devise linear models with random walks, taking both intra- and inter-session correlations into account.

%% file: sec-model.tex
\section{S-Walk: The Proposed Model}\label{sec:model}

\input{Figures/Fig3_swalk}

In this section, we propose a novel session-based recommendation with a random walk, namely \emph{S-Walk}. While existing random-walk-based recommender models~\cite{YildirimK08,JamaliE09,CooperLRS14,ChristoffelPNB15,EksombatchaiJLL18,NikolakopoulosK19,NikolakopoulosK20} are based on a user-item bipartite graph, it is non-trivial to adopt them for session-based recommendations. Since user information is unavailable, we rely only on item information to learn underlying patterns. Furthermore, the session-item matrix is extremely sparse.

To address these issues, we first present the overall architecture of S-Walk using \emph{global item graphs} (Section~\ref{sec:architecture}). Intuitively, walking on a global item graph can describe inter-session relationships among items because the walker can move from an item of the current session to the items of other sessions. Then, we develop two linear models to build a transition matrix and a teleportation matrix, used in S-Walk (Sections~\ref{sec:transtion_matrix}--\ref{sec:teleporation_matrix}). Note that these linear models can be replaced by others as long as they are efficient and scalable. Notably, our linear models satisfy the desirable condition for efficiency and scalability. Finally, we explain model training and inference of S-Walk (Section~\ref{sec:training}).

\subsection{Model Architecture}
\label{sec:architecture}

Figure~\ref{fig:swalk} overviews the S-Walk model. Given a session-item interaction matrix $\mathbf{X}$, we first design two models for item transition and teleportation to capture different characteristics of sessions (the blue and orange box in Figure~\ref{fig:swalk}, respectively). We then constitute a final global item graph using \emph{random walk with restart (RWR)}.

Specifically, each model produces its own relevance matrix over the transition graph $\mathcal{G}_{\mathbf{R}} = (\mathcal{I}, \mathcal{E}_{\mathbf{R}})$ and the teleportation graph $\mathcal{G}_{\mathbf{T}} = (\mathcal{I}, \mathcal{E}_{\mathbf{T}})$, where each node corresponds to an item and an edge indicates the relevance between a pair of items. The \emph{transition matrix} $\mathbf{R}$ is the adjacency matrix of $\mathcal{G}_{\mathbf{R}}$ which encodes sequential dependency and repeated item consumption in a session. On the other hand, the \emph{teleportation matrix} $\mathbf{T}$ is the adjacency matrix of $\mathcal{G}_{\mathbf{T}}$, which captures item consistency in the session. Introducing the two matrices captures different intra-session relationships among items, but they do not address the inter-session relationship.

By adopting the RWR using the two graphs, where a random walker jumps from one node to another or restarts on an arbitrary node regardless of her current position, we intend to consider the inter-session relationship, capturing high-order relationships among items, \ie, multi-hop connections on the item graphs. 
Conceptually, the RWR on the two item graphs, $\mathcal{G}_R$ and $\mathcal{G}_T$, can be thought of as tossing a biased coin that yields the head with probability $\alpha$:
\begin{enumerate}
    \item If the coin is head (with a probability of $\alpha$), the walker moves to one of the items adjacent to the current item through the transition matrix $\mathbf{R} \in \mathbb{R}^{n \times n}$.
    
    \item If the coin is tail (with a probability of $1 - \alpha$), the walker restarts on one of the items adjacent to the start item through the teleportation matrix $\mathbf{T} \in \mathbb{R}^{n \times n}$.
\end{enumerate}

The random walk using these two matrices is a stochastic process, which can also be seen as a Markov chain on items over a homogeneous discrete time. Formally, we formulate the RWR as follows:
\begin{equation}\label{eq:rwr}
\mathbf{x}_{(k+1)} = \alpha \mathbf{x}_{(k)} \mathbf{R} + (1 - \alpha) \mathbf{x}_{(0)} \mathbf{T}, \ \ \text{where} \ \ k = 0, \dots, \infty.
\end{equation}
Here, $\alpha$ is the damping factor that controls the proportion of the random walk and the restart. $\mathbf{x}^{\top}_{(0)} \in \mathbb{R}^{n}$ is the initial item vector and $\mathbf{x}^{\top}_{(k)} \in \mathbb{R}^{n}$ is the updated proximity score for items after the $k$-th step. As the $k$ increases, $\mathbf{x}_{(k)}$ converges to limited distribution. Through the RWR, we obtain stationary probabilities that the random walker lands on each node, expressed as the green graph in Figure~\ref{fig:swalk}. Finally, we generate the recommendation list using the final graph $\mathcal{G}_{\mathbf{M}} = (\mathcal{I}, \mathcal{E}_{\mathbf{M}})$.

In this process, we devise linear models for the two models, taking the following advantages: (i) they achieve comparable performance without complicated tuning, and (ii) training and inference are much faster than DNN-based session recommender models~\cite{HidasiKBT15,HidasiQKT16,HidasiK18,LiRCRLM17,LiuZMZ18,WuT0WXT19,XuZLSXZFZ19,GuptaGMVS19,PanCLR20a,PanCCCR20}.

\subsection{Item Transition Model}
\label{sec:transtion_matrix}

Fist, we develop a linear transition model to build the item transition matrix $\mathbf{R}$. As a natural way of representing the transition of item sequences, we introduce \emph{partial session representations}. A session $s$ is divided into two sub-sessions, \emph{past} and \emph{future}, according to each time step $t = 1, ..., |s|$. The past partial session consists of items consumed before the $t$-th item, \ie, $\mathbf{s}_{1:t-1} = \{s_1, \ldots, s_{t-1}\}$. The future partial session consists of items consumed at or after the $t$-th item, \ie, $\mathbf{s}_{t:|s|} = \{s_t, \dots, s_{|s|}\}$. For each time $t = 2, ..., |s|$, we produce $|s| - 1$ past and future partial session pairs. By stacking $|s| - 1$ pairs for all $s \in \mathcal{S}$, we build two matrices, the past session matrix $\mathbf{Y} \in \mathbb{R}^{m' \times n}$ and future session matrix $\mathbf{Z} \in \mathbb{R}^{m' \times n}$, where $m'$ is the number of all partial sessions, \ie, $m' = \sum_{i=1}^{m}{(|s^{(i)}| - 1})$. Finally, the item transition matrix $\mathbf{R}$ is learned with two matrices $\mathbf{Y}$ and $\mathbf{Z}$ according to the partial session representation.

To represent the temporal proximity of items, we adjust the weights of items in $\mathbf{Y}$ and $\mathbf{Z}$. As adopted in~\cite{GargGMVS19, ChoiKLSL21SLIST}, we consider the position gap between two items as the weight of items within a session.
\begin{equation}
  \label{eq:weight_item}
  w_\text{pos}(i, j, s) = \text{exp}\left(-\frac{|p(i, s) - p(j, s)|}{\delta_\text{pos}} \right),
\end{equation}
where $\delta_\text{pos}$ is the hyper-parameter that controls the position decay in partial sessions, and $p(i, s)$ is the position of item $i$ in the session $s$. In this way, we decay the relevance between items $i$ and $j$ as they get farther away.

Formally, the item transition model is formulated as
\begin{equation}\label{eq:transition}
  \argmin_{\mathbf{B}^\text{tran}} \left\| \mathbf{Z} - \mathbf{Y} \cdot \mathbf{B}^\text{tran} \right\|_F^2 + \lambda \|\mathbf{B}^\text{tran}\|_F^2,
\end{equation}
where $\mathbf{B}^\text{tran}$ is the item-item relevance matrix learned from the sequential dependency between $\mathbf{Y}$ and $\mathbf{Z}$. As $\mathbf{Y} \ne \mathbf{Z}$, we can naturally avoid the trivial solution $\mathbf{B}^\text{tran} = \mathbf{I}$. Unlike Eq.~\eqref{eq:ease_loss}, we can remove the zero-diagonal constraint in $\mathbf{B}^\text{tran}$. The closed-form solution is given by
\begin{equation}
  \label{eq:trans_solution}
  \hat{\mathbf{B}}^\text{tran} = \hat{\mathbf{P}}' \cdot (\mathbf{Y}^{\top} \mathbf{Z}),
\end{equation}
where $\hat{\mathbf{P}}' = (\mathbf{Y}^{\top} \mathbf{Y} + \lambda \mathbf{I})^{-1} \in \mathbb{R}^n$. The computational complexity is independent of the number of users, as shown in Eq.~\eqref{eq:ease_solution}. (See Appendix A for a detailed derivation of our solution in the supplementary material.)

To utilize the item transition matrix in random walks, each element should be the transition probability from one node to another. That is, every element is non-negative and sums to 1. However, $\hat{\mathbf{B}}^\text{tran}$ is not normally a probability matrix. To satisfy the non-negative constraint, we first replace all negative values in $\hat{\mathbf{B}}^\text{tran}$ with zero\footnote{As the negative values in $\hat{\mathbf{B}}^\text{tran}$ mean negative correlations among items, they are less important than positive correlations. In our empirical study, it is observed that this conversion does not harm the accuracy of the linear model.}, denoting $\hat{\mathbf{B}}^\text{tran}_{\ge 0}$.
We then normalize $\hat{\mathbf{B}}^\text{tran}_{\ge 0}$ as follows:
\begin{equation}
  \mathbf{R} = \texttt{diagMat}(\hat{\mathbf{B}}^{\text{tran}}_{\ge 0}\mathbf{1})^{-1} \hat{\mathbf{B}}^{\text{tran}}_{\ge 0}.
\end{equation}
Here, $\mathbf{R}$ is the item transition probability matrix, where each row is normalized, and $\mathbf{1}$ is a column vector of length $n$ filled with one.

\subsection{Item Teleportation Model}
\label{sec:teleporation_matrix}

The item teleportation matrix is designed to capture the item consistency within a session. For this reason, we focus on modeling co-occurrence between items. A session is treated as a set of items $\mathbf{s} = \{s_1, s_2, \ldots, s_{|s|}\}$, ignoring the order of items. By stacking $m$ sessions, we build a binary session-item matrix $\mathbf{X}$. Note that the repeated items in the session are treated as a single item.

Given the matrix $\mathbf{X}$, we devise a linear teleportation model. It is formulated with the same input and output matrix as used in the existing linear models~\cite{NingK11, Steck19b}. Meanwhile, we relax the zero-diagonal constraint for $\mathbf{B}$ to handle repeated item consumption, as discussed in~\cite{ChoiKLSL21SLIST}. When the diagonal element of $\mathbf{B}$ is loosely penalized, it allows us to predict the same items as the next item repeatedly.
\begin{equation}
  \label{eq:tele_loss}
  \argmin_{\mathbf{B}^\text{tele}} \left\| ( \mathbf{X} - \mathbf{X} \cdot \mathbf{B}^\text{tele} )\right\|_F^2 + \lambda \|\mathbf{B}^\text{tele}\|_F^2,
\ \ \text{s.t.} \ \ \texttt{diag}(\mathbf{B}^\text{tele}) \le \xi,
\end{equation}
where $\mathbf{B}^\text{tele}$ is the item-item relevance matrix for item consistency, and $\xi$ is the hyper-parameter to control diagonal constraints. When $\xi = 0$, it is equivalent to the zero-diagonal constraint for $\mathbf{B}^\text{tele}$. When $\xi = \infty$, there is no constraint on the diagonal elements of $\mathbf{B}^\text{tele}$. Note that the objective function of EASE$^{R}$~\cite{Steck19b} is a special case of Eq.~\eqref{eq:tele_loss}, where $\mathbf{B}^\text{tele}$ with $\xi=0$.

We can obtain the closed-form solution of $\mathbf{B}^\text{tele}$ as follows:
\begin{align}
    \hat{\mathbf{B}}^\text{tele} = \mathbf{I} - \hat{\mathbf{P}} \cdot  \texttt{diagMat}(\mathbf{\gamma}),
    \ \ \mathbf{\gamma}_{j} = 
    \begin{cases}
        \ \lambda & \text{if} \ 1 - \lambda P_{jj} \le \xi, \\
        \ \frac{1-\xi}{P_{jj}} & \text{otherwise},
    \end{cases}
  \label{eq:tele_solution}
\end{align}
where $\hat{\mathbf{P}} = (\mathbf{X}^{\top} \mathbf{X} + \lambda \mathbf{I})^{-1}$, and $\mathbf{\gamma} \in \mathbb{R}^n$ is the vector used to check the diagonal constraint of $\mathbf{B}^\text{tele}$. Because of the inequality condition for the diagonal elements in $\mathbf{B}^\text{tele}$, $\mathbf{\gamma}_{j}$ is determined by $(1 - \lambda P_{jj})$. The closed-form solution may depend on $\xi$. If $\xi = 0$, $\mathbf{\gamma}_{j}$ is equal to ${1} / P_{jj}$, corresponding to $(1 \oslash \texttt{diag}(\hat{\mathbf{P}}))$. If $\xi = \infty$, the solution becomes $\hat{\mathbf{B}}^{\text{tele}}=\mathbf{I}-\lambda\hat{\mathbf{P}}$.

Similar to $\hat{\mathbf{B}}^\text{tran}$, the solution $\hat{\mathbf{B}}^\text{tele}$ does not satisfy the non-negativity and normalization conditions to be a probability matrix. We compute the item teleportation probability matrix $\mathbf{T}$ by replacing the negative values with zero and normalizing $\hat{\mathbf{B}}^\text{tele}_{\ge 0}$:
\begin{equation}\label{eq:tele_matrix}
  \mathbf{T} = \beta \left(\texttt{diagMat}(\hat{\mathbf{B}}^\text{tele}_{\ge 0}\mathbf{1})^{-1} \hat{\mathbf{B}}^\text{tele}_{\ge 0}\right) + \left(1 - \beta \right)\mathbf{I},
\end{equation}
where $\beta$ is a hyper-parameter that controls the importance of the self-loop to guarantee the convergence of random walks. (See Appendix~\ref{sec:analysis} for detailed analysis in the supplementary material.)

\subsection{Random-walk Training and Inference}\label{sec:training}

\textbf{Training.} To compute the stationary distribution of S-Walk, we utilize the power method~\cite{AmyNCarlD06pagerank}. Each proximity score corresponds to the limiting distribution using Eq.~\eqref{eq:rwr}:
\begin{equation} 
\begin{split}
\mathbf{x}_{(\infty)} 
& = \alpha^{\infty} \mathbf{x}_{(0)} \mathbf{R}^{\infty} + \sum_{k=0}^{\infty}{\alpha}^{k}{(1-\alpha)} \mathbf{x}_{(0)} \mathbf{T}\mathbf{R}^{k} \\
& \approx \mathbf{x}_{(0)} {\sum_{k=0}^{\infty}{\alpha}^{k}{(1-\alpha)}\mathbf{T}\mathbf{R}^{k}} = \mathbf{x}_{(0)} \mathbf{M}. 
\end{split}
\label{eq:final_RWR_swalk}
\end{equation}

\noindent
Here, $\mathbf{M} = {\sum_{k=0}^{\infty}{\alpha}^{k}{(1-\alpha)}\mathbf{T}\mathbf{R}^{k}}$ is the trained item-item matrix by S-Walk. (Appendix~\ref{sec:algorithm} provides the detailed pseudo-code for computing the proximity score using the power method.)

\vspace{1mm}
\noindent
\textbf{Inference.} Given a new session $s_\text{new}$, we represent a session vector $\mathbf{x}_\text{new}$ and compute the proximity score using $\mathbf{M}$. The proximity score for predicting the next item is finally given by $\mathbf{x}_{\text{new}} \mathbf{M}$. In this process, we decay the importance of items in $\mathbf{x}_\text{new}$ to rely more on recent consumption, similar to Eq.~\eqref{eq:weight_item}:
\begin{equation}
  \label{eq:weight_inference}
  w_\text{inf}(i, s_{\text{new}}) = \text{exp}\left(-\frac{|s_{\text{new}}| - p(i, s_{\text{new}})}{\delta_\text{inf}} \right),
\end{equation}
where $\delta_\text{inf}$ is the hyper-parameter used to control the item weight decay, and $|s_{\text{new}}|$ is the length of the session $s_{\text{new}}$.

%% file: Figures/Fig3_swalk.tex
\begin{figure*}
\includegraphics[width=0.95\linewidth]{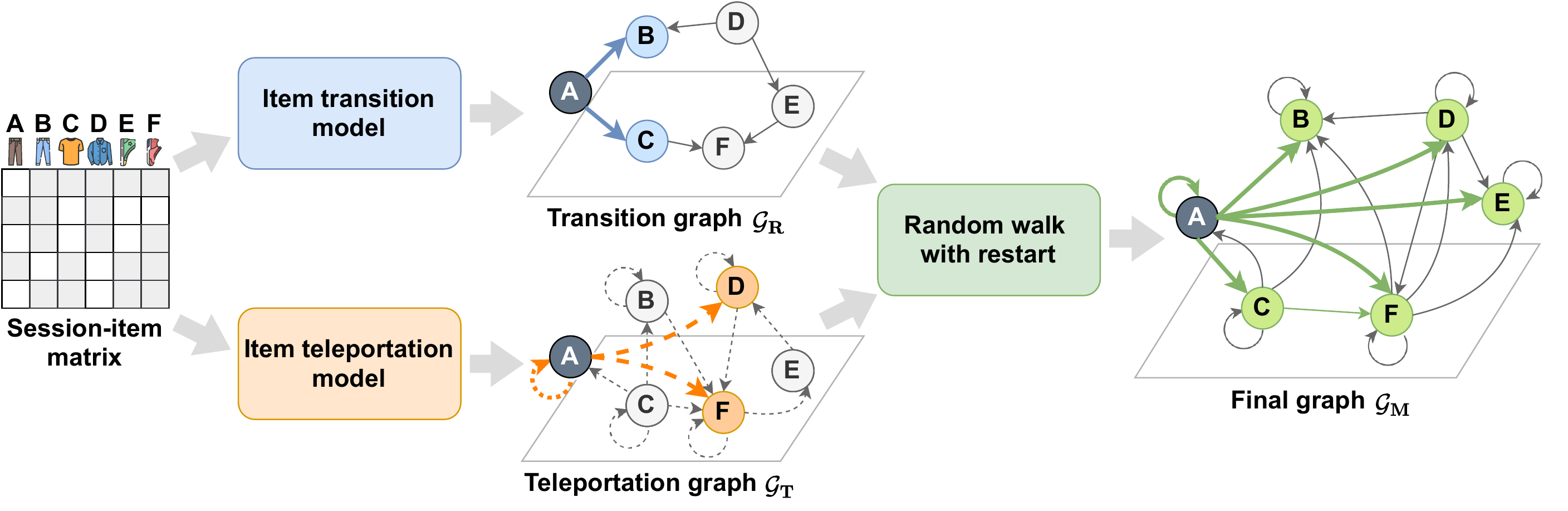}
\caption{The overall architecture of S-Walk. Given a session-item matrix, two different linear models build the transition graph and the teleportation graph by adjacency matrices $\mathbf{R}$ and $\mathbf{T}$, using Eq.~\eqref{eq:trans_solution} and Eq.~\eqref{eq:tele_solution}, respectively. Then, in Eq.~\eqref{eq:final_RWR_swalk}, the random walk with restart is used to build the final graph with adjacency matrix $\mathbf{M}$, capturing high-order relationships.}\label{fig:swalk}
\vskip -0.1in
\end{figure*}

%% file: sec-experiments.tex
\section{Experimental Setup}
\label{sec:setup}

\textbf{Benchmark datasets}. We use four public datasets collected from e-commerce and music streaming services: YooChoose\footnote{\url{https://www.kaggle.com/chadgostopp/recsys-challenge-2015}} (YC), Diginetica\footnote{\url{https://competitions.codalab.org/competitions/11161}} (DIGI), RetailRocket\footnote{\url{https://www.kaggle.com/retailrocket/ecommerce-dataset}} (RR), and NowPlaying\footnote{\url{https://drive.google.com/drive/folders/1ritDnO_Zc6DFEU6UND9C8VCisT0ETVp5}} (NOWP). Following existing studies~\cite{HidasiK18,LiRCRLM17,LiuZMZ18,WuT0WXT19}, we use single training-test split (single-split) datasets (\ie, YC-1/4, DIGI1). To minimize the risk of random effects, we also split the datasets (\ie, YC5, DIGI5, RR, NOWP) into five folds (five-split) which are contiguous in time, as used in recent extensive evaluation~\cite{LudewigJ18,LudewigMLJ19a,LudewigMLJ19b}. Following the convention~\cite{LudewigJ18,LudewigMLJ19a,LudewigMLJ19b}, we discard the sessions with only one interaction and items that occur less than five times. Also, we split the training and test datasets chronologically and use a portion of the training set for validation, such that the validation set size is equal to that of the test dataset. (See Appendix~\ref{sec:dataset} for detailed statistics of all benchmark datasets.)

\vspace{1mm}
\noindent
\textbf{Competing models}. We compare our model to nine competing models. Among the non-neural models, we compare to AR~\cite{AgrawalIS93}, SR~\cite{KamehkhoshJL17}, STAN~\cite{GargGMVS19}, and SLIST~\cite{ChoiKLSL21SLIST}. AR~\cite{AgrawalIS93} is a simple model using association rules, and SR~\cite{KamehkhoshJL17} is a Markov chain model together with association rules. STAN~\cite{GargGMVS19} is an improved model of SKNN~\cite{JannachL17}, a session-based kNN algorithm. SLIST~\cite{ChoiKLSL21SLIST} is a linear model designed for SR. Among the neural models, we compare to NARM~\cite{LiRCRLM17}, STAMP~\cite{LiuZMZ18}, SR-GNN~\cite{WuT0WXT19}, NISER+~\cite{GuptaGMVS19}, and GCE-GNN~\cite{WangWCLMQ20}. Notably, SR-GNN~\cite{WuT0WXT19} employs GNNs to consider complex sequential dependency of items. To overcome overfitting and popularity bias, NISER+~\cite{GuptaGMVS19} uses normalized items and session embeddings on top of SR-GNN. Lastly, GCE-GNN~\cite{WangWCLMQ20} considers inter-session relationships by modeling item transitions over all sessions. (See Appendix~\ref{sec:app_reproducibility} for the implementation details of baselines.) All source codes are available at https://github.com/jin530/SWalk.

\vspace{1mm}
\noindent
\textbf{Evaluation protocol and metrics}. As in the common protocol~\cite{HidasiQKT16,LiRCRLM17,WuT0WXT19}, we use the \emph{iterative revealing scheme}, which iteratively exposes an item from a session to the model, to reflect sequential user behavior throughout the entire session. We adopt several metrics to handle two scenarios: (i) To evaluate the next single item, we use \emph{Hit Rate (HR)} and \emph{Mean Reciprocal Rank (MRR)}, which have been widely used in existing studies~\cite{HidasiKBT15,LiuZMZ18,WuT0WXT19}. HR and MRR measure the next item's existence and rank in the recommendation list, respectively. (ii) To evaluate all subsequent items predicted, we use two common IR measures, \emph{Recall (R@k)} and \emph{Mean Average Precision (MAP@k)}, which measure the subsequent items' existence and rank in the recommendation list, respectively. We use $k = \{5, 10, 20, 50, 100\}$, where $k$ = 20 is the default.

%% file: sec-results.tex
\section{Evaluation Results}\label{sec:results}

We evaluate the accuracy and efficiency of S-Walk by comparing it with competing models. Based on thorough and extensive evaluations, we make the following conclusions.

\begin{itemize}[leftmargin=5mm]
\item S-Walk exhibits \textbf{state-of-the-art performance} on multiple datasets.
For R@20, MAP@20, and HR@20, S-Walk consistently outperforms the other models, up to 3.16\%, 4.11\%, and 3.65\%, respectively (Section~\ref{sec:exp_overall}).

\item The inference of S-Walk is up to \textbf{8.9 times faster} than other DNN-based models. Surprisingly, S-Walk can be \textbf{compressed} (sparsified by zeroing-out entries via thresholding), \textbf{without sacrificing its accuracy} (Section~\ref{sec:exp_time}).

\item For \textbf{long sessions}, where user intent is more difficult to capture, S-Walk significantly outperforms existing models. (Section~\ref{sec:exp_length}).

\item S-Walk \textbf{converges within 3--5 steps} and is superior to the k-step method because it utilizes the teleportation model for restarts (Section~\ref{sec:exp_ablation}).
\end{itemize}

\subsection{Evaluation of Accuracy}
\label{sec:exp_overall}

\input{Tables/tab2_accuracy}

\input{Tables/tab3_scalability}

Table~\ref{tab:overallresult} reports the accuracy of S-Walk and other competitors. It is found that no single model shows the best performance over all the datasets, as reported in existing studies~\cite{LudewigJ18,LudewigMLJ19a,LudewigMLJ19b}. Remarkably, the existing models have different tendencies for single-split and five-split datasets; while most neural models surpass non-neural models on the single-split datasets, their performances are degraded on the five-split datasets. We conjecture that the parameters of the neural models reported in existing studies are mostly biased for optimizing the single-split datasets. For five-split datasets, it is also observed that the variance in the neural models is larger than that of non-neural models. Based on these observations, it is challenging for one model to consistently achieve outstanding performance on all the datasets.

Nevertheless, S-Walk shows competitive or state-of-the-art performances. Notably, the gain of S-Walk is up to 4.08\% and 8.15\% in R@20 and MAP@20 over the best competing model. For all five-split datasets, S-Walk consistently surpasses the existing models for HR@20, R@20, and MAP@20. These empirical results indicate that S-Walk captures various intra- and inter-correlation among items without being biased to specific datasets.

Compared to the other metrics, S-walk is slightly lower on MRR@20, particularly on YC and NOWP datasets. Based on the gap between SR~\cite{KamehkhoshJL17} and STAN~\cite{GargGMVS19}, we observe that the MRR@20 scores on these datasets are mostly affected by intra-session relationships. Thus, we address this by taking fewer random walks on these datasets; \eg, S-Walk$_{(1)}$ considers mostly intra-session relationships, confirming superior performance in MRR@20 as well.

\subsection{Evaluation of Scalability}
\label{sec:exp_time}

Table~\ref{tab:time} compares the inference time between S-Walk and the other best models on several datasets. Whereas the computational cost of S-Walk is proportional only to the number of items, that of neural models also depends on the number of layers and their dimensions. Although S-Walk runs on CPU and other models run on GPU, S-Walk shows about five times faster inference time owing to its simpler structure, even with better accuracy. This property is highly desirable for deploying S-Walk for real-world applications.

We also attempt \emph{model pruning} for S-Walk. As a simple strategy, we adopt magnitude pruning~\cite{FrankleC19}, \ie, the model is globally sparsified by zeroing out parameter values with the smallest magnitude. For instance, 100$\times$ compression means that we retain only 1\% non-zero entries in $\mathbf{M}$ with the highest magnitude while zeroing out the remaining 99\%. Surprisingly, S-Walk preserves the accuracy with highly extreme (100$\times$) compression ratios, as depicted in Figure~\ref{fig:model_compression}. Random pruning is a simple baseline, \ie, the parameters are randomly removed under the given compression ratio. This result indicates that the learned relationship between items is disentangled with the locality, as observed in existing studies~\cite{Lee2013iclr,Lee2013llorma,Lee2014lcr,LeeKLSB16,Choi2021loca}. Owing to this valuable property, the compressed S-Walk is memory-efficient, suitable for low-resource devices, \eg, mobile and embedded applications.

\input{Figures/Fig3_pruning}
\input{Figures/Fig4_session_length_baseline}

\subsection{Effect of Various Session Data}\label{sec:exp_length}

\input{Figures/Fig5_cutoff}

To further investigate the accuracy gains of S-Walk we carefully design case studies and examine the effects of session length and retrieval size. (See Appendix~\ref{sec:app_additionalresults} for the effect of the data size on S-Walk and competing models.)

To observe how the session length affects the performance, we compare it with other best models (\ie, STAN~\cite{GargGMVS19}, SR-GNN~\cite{WuT0WXT19} and NISER+~\cite{GuptaGMVS19}) by categorizing the entire sessions into two groups: short ($\le 5$ items) and long sessions ($> 5$ items). Note that the ratio of short sessions is 77.2\% (DIGI5) and 79.2\% (RR), respectively. Figure~\ref{fig:length_baseline} indicates that long sessions are much more challenging to predict users' hidden intents effectively. Nonetheless, S-Walk significantly outperforms NISER+~\cite{GuptaGMVS19} in long sessions by 6.2\%--20.2\% and 5.8\%--21.0\% on R@20 and MAP@20, respectively. Based on this result, we confirm that S-Walk effectively captures complicated item patterns, further improving the accuracy of longer sessions.

Figure~\ref{fig:cutoff} depicts the comparison results between S-Walk and the competitive models for various numbers of recommended items. For all cut-off sizes, we observe that S-Walk consistently surpasses NISER+ by 5.2\%--16.9\% and 5.4\%--17.2\% gains on Recall and MAP, respectively. In this sense, S-Walk can expand item coverage by increasing the number of steps through the random walk process in the item graph.


\subsection{Ablation Study}\label{sec:exp_ablation}

\input{Tables/tab4_ablation}

We analyze the effect of each component, \ie, the transition model and the teleportation model, in S-Walk. As shown in Table~\ref{tab:ablation}, the complete S-Walk shows the best performance, compared to using SR~\cite{KamehkhoshJL17} for the transition model. For the teleportation model, AR~\cite{AgrawalIS93} shows worse accuracy than the identity matrix in the YC-1/4 dataset, where AR~\cite{AgrawalIS93} shows the worst performance. This implies that incorrect teleportation may hinder random walks.

Figure~\ref{fig:combine_alpha} depicts the effect of the damping factor $\alpha$ in S-Walk, controlling the ratio of walks and restarts. With $\alpha = 0.7$, S-Walk shows the best performance, implying that the transition model is more dominant than the teleportation model. Finally, with various numbers of random walk steps, we compare S-walk with the $k$-step method, which uses the $k$-step landing probability on the transition graph. (i) Without restart using the teleportation graph, performance significantly degrades, and (ii) when $k \ge 2$ in $k$-step, the accuracy continues to decrease. (iii) S-Walk usually converges within 3--5 steps, achieving the best performance on both datasets.

\input{Figures/Fig7_alpha_hyperparameter}

%% file: Tables/tab2_accuracy.tex
\begin{table*}[t] \small
\caption{Accuracy comparison of S-Walk and competing models, following experimental setup in \cite{LudewigMLJ19a, LudewigMLJ19b}. Gains indicate how much better the best proposed model is than the best baseline model. S-Walk$_{(1)}$ is a variant of S-Walk trained only up to the first step $\textbf{M}_{(1)}$. The best model is marked in \com{\textbf{bold}} and the second best model is \fcom{\underline{underlined}}.}
\vskip -0.1in
\label{tab:overallresult}
\begin{center}
\begin{tabular}{c|c|cccc|ccccc|cc|c}
\toprule
\multirow{2}{*}{Dataset} & \multirow{2}{*}{Metric} & \multicolumn{4}{c|}{Non-Neural Models}        & \multicolumn{5}{c|}{Neural Models}                                    & \multicolumn{2}{c|}{Ours}         & \multirow{2}{*}{Gain(\%)} \\
                         &                         & AR     & SR     & STAN         & SLIST        & NARM   & STAMP  & SR-GNN          & NISER+          & GCE-GNN         & S-Walk$_{(1)}$  & S-Walk          &                           \\ \midrule
\multirow{4}{*}{DIGI5}   & R@20                    & 0.2872 & 0.2517 & 0.3720       & 0.3803       & 0.3254 & 0.3040 & 0.3232          & 0.3727          & \fcom{\underline{0.3927}} & 0.3761          & \com{\textbf{0.3995}} & 1.73                      \\
                         & MAP@20                  & 0.0189 & 0.0164 & 0.0252       & 0.0259       & 0.0218 & 0.0201 & 0.0217          & 0.0259          & \fcom{\underline{0.0271}} & 0.0254          & \com{\textbf{0.0277}} & 2.21                      \\
                         & HR@20                   & 0.3720 & 0.3277 & 0.4800       & 0.4915       & 0.4188 & 0.3917 & 0.4158          & 0.4785          & \fcom{\underline{0.5086}} & 0.4873          & \com{\textbf{0.5115}} & 0.57                      \\
                         & MRR@20                  & 0.1280 & 0.1216 & \fcom{\underline{0.1828}} & 0.1809       & 0.1392 & 0.1314 & 0.1436          & 0.1656          & 0.1803                & 0.1732          & \com{\textbf{0.1890}} & 3.39                \\ \hline
\multirow{4}{*}{RR}      & R@20                    & 0.3533 & 0.3359 & 0.4748       & 0.4724       & 0.4526 & 0.3917 & 0.4438          & 0.4630          & \fcom{\underline{0.4841}}   & 0.4810          & \com{\textbf{0.4994}} & 3.16                      \\
                         & MAP@20                  & 0.0205 & 0.0194 & 0.0285       & 0.0282       & 0.0270 & 0.0227 & 0.0264          & 0.0278          & \fcom{\underline{0.0292}}    & 0.0290          & \com{\textbf{0.0304}} & 4.11                      \\
                         & HR@20                   & 0.4367 & 0.4174 & 0.5938       & 0.5877       & 0.5549 & 0.4620 & 0.5433          & 0.5651          & 0.6007          & \fcom{\underline{0.6019}}    & \com{\textbf{0.6226}} & 3.65                      \\
                         & MRR@20                  & 0.2407 & 0.2453 & \fcom{\underline{0.3638}} & 0.3131       & 0.3196 & 0.2527 & 0.3066          & 0.3171          & 0.3362          & 0.3409          & \com{\textbf{0.3645}} & 0.19                      \\ \hline
\multirow{4}{*}{YC5}     & R@20                    & 0.4760 & 0.4853 & 0.4986       & 0.5122       & 0.5109 & 0.4979 & 0.5060          & \fcom{\underline{0.5146}}   & 0.4972          & 0.5096          & \com{\textbf{0.5189}} & 0.85                      \\
                         & MAP@20                  & 0.0325 & 0.0332 & 0.0342       & 0.0357       & 0.0357 & 0.0344 & 0.0351          & \fcom{\underline{0.0363}}    & 0.0348          & 0.0355          & \com{\textbf{0.0365}} & 0.55                      \\
                         & HR@20                   & 0.6361 & 0.6506 & 0.6656       & \fcom{\underline{0.6867}} & 0.6751 & 0.6654 & 0.6713          & 0.6858          & 0.6650          & 0.6834          & \com{\textbf{0.6906}} & 0.57                      \\
                         & MRR@20                  & 0.2894 & 0.3010 & 0.2933       & 0.3097       & 0.3047 & 0.3033 & \com{\textbf{0.3142}} & \fcom{\underline{0.3130}}    & 0.2939          & 0.3074          & 0.2892          & -2.17                     \\ \hline
\multirow{4}{*}{NOWP}    & R@20                    & 0.1544 & 0.1366 & 0.1696       & \fcom{\underline{0.1840}} & 0.1274 & 0.1253 & 0.1400          & 0.1493          & 0.1504          & 0.1837          & \com{\textbf{0.1915}} & 4.08                      \\
                         & MAP@20                  & 0.0166 & 0.0133 & 0.0175       & \fcom{\underline{0.0184}} & 0.0118 & 0.0113 & 0.0125          & 0.0143          & 0.0143          & 0.0182          & \com{\textbf{0.0199}} & 8.15                      \\
                         & HR@20                   & 0.2076 & 0.2002 & 0.2414       & \fcom{\underline{0.2689}} & 0.1849 & 0.1915 & 0.2113          & 0.2196          & 0.2122          & 0.2678          & \com{\textbf{0.2693}} & 0.15                      \\
                         & MRR@20                  & 0.0710 & 0.1052 & 0.0871       & \fcom{\underline{0.1137}} & 0.0894 & 0.0882 & 0.0935          & 0.0944          & 0.0720          & \com{\textbf{0.1158}} & 0.0866          & 1.85                      \\ \hline
\multirow{4}{*}{DIGI1}   & R@20                    & 0.3401 & 0.3164 & 0.3965       & 0.4091       & 0.3890 & 0.3663 & 0.3811          & \com{\textbf{0.4202}} & 0.4169          & 0.4012          & \fcom{\underline{0.4201}}    & -0.02                     \\
                         & MAP@20                  & 0.0231 & 0.0212 & 0.0275       & 0.0286       & 0.0270 & 0.0252 & 0.0264          & \com{\textbf{0.0299}} & 0.0295          & 0.0278          & \fcom{\underline{0.0297}}    & -0.67                     \\
                         & HR@20                   & 0.4394 & 0.4085 & 0.5121       & 0.5291       & 0.4979 & 0.4690 & 0.4896          & \fcom{\underline{0.5397}}    & 0.5362          & 0.5188          & \com{\textbf{0.5420}} & 0.43                      \\
                         & MRR@20                  & 0.1465 & 0.1431 & 0.1851       & 0.1886       & 0.1585 & 0.1499 & 0.1654          & 0.1856          & \fcom{\underline{0.1907}}    & 0.1869          & \com{\textbf{0.1927}} & 1.05                      \\ \hline
\multirow{4}{*}{YC-1/4}  & R@20                    & 0.4781 & 0.4851 & 0.4952       & 0.5130       & 0.5097 & 0.5008 & 0.5095          & \fcom{\underline{0.5164}}   & 0.5030          & 0.5103          & \com{\textbf{0.5213}} & 0.95                      \\
                         & MAP@20                  & 0.0363 & 0.0364 & 0.0373       & 0.0393       & 0.0395 & 0.0385 & 0.0393          & \fcom{\underline{0.0402}}    & 0.0388          & 0.0390          & \com{\textbf{0.0404}} & 0.50                      \\
                         & HR@20                   & 0.6697 & 0.6850 & 0.6846       & 0.7175       & 0.7079 & 0.7021 & 0.7118          & \fcom{\underline{0.7182}}    & 0.7036          & 0.7145          & \com{\textbf{0.7204}} & 0.31                      \\
                         & MRR@20                  & 0.2880 & 0.3053 & 0.2829       & \fcom{\underline{0.3161}} & 0.2996 & 0.3066 & \com{\textbf{0.3180}} & \fcom{\underline{0.3161}}    & 0.3027          & 0.3137          & 0.2867          & -1.35                     \\ \bottomrule
\end{tabular}
\end{center}
\vskip -0.1in
\end{table*}

%% file: Tables/tab3_scalability.tex
\begin{table}[t] \small
\caption{The number of floating point operations and runtime (in seconds) for inference of S-Walk and DNN-based models. Gain indicate how fast S-Walk is compared to GCE-GNN~\cite{WangWCLMQ20}. Runtime for DNN models was measured on GPU, while that for S-Walk on CPU.}\label{tab:time}
\vspace{-0.1in}
\begin{tabular}{l|cc|cc|cc}
\toprule
\multirow{2}{*}{Models} & \multicolumn{2}{c|}{YC1/4} & \multicolumn{2}{c|}{DIGI5} & \multicolumn{2}{c}{RR} \\
                        & GFLOPs        & Time       & GFLOPs        & Time       & GFLOPs      & Time      \\ \midrule
SR-GNN                  & 1282.8        & 70.8       & 765.4         & 49.2       & 247.2       & 12.7      \\
NISER+                  & 2605.8        & 87.1       & 1551.0        & 59.7       & 501.8       & 15.7      \\
GCE-GNN                 & 51094.8       & 108.8      & 10445.9       & 74.0       & 9446.0      & 19.8      \\
S-Walk                  & 11.0          & 20.5       & 4.9           & 8.3        & 2.3         & 5.2       \\ \midrule
Gain                    & 4632.3x        & 5.3x        & 2131.3x        & 8.9x        & 4133.2x      & 3.8x       \\ \bottomrule
\end{tabular}
\vskip -0.1in
\end{table}

%% file: Figures/Fig3_pruning.tex
\begin{figure}
\centering

\begin{tabular}{cc}
\includegraphics[width=0.20\textwidth]{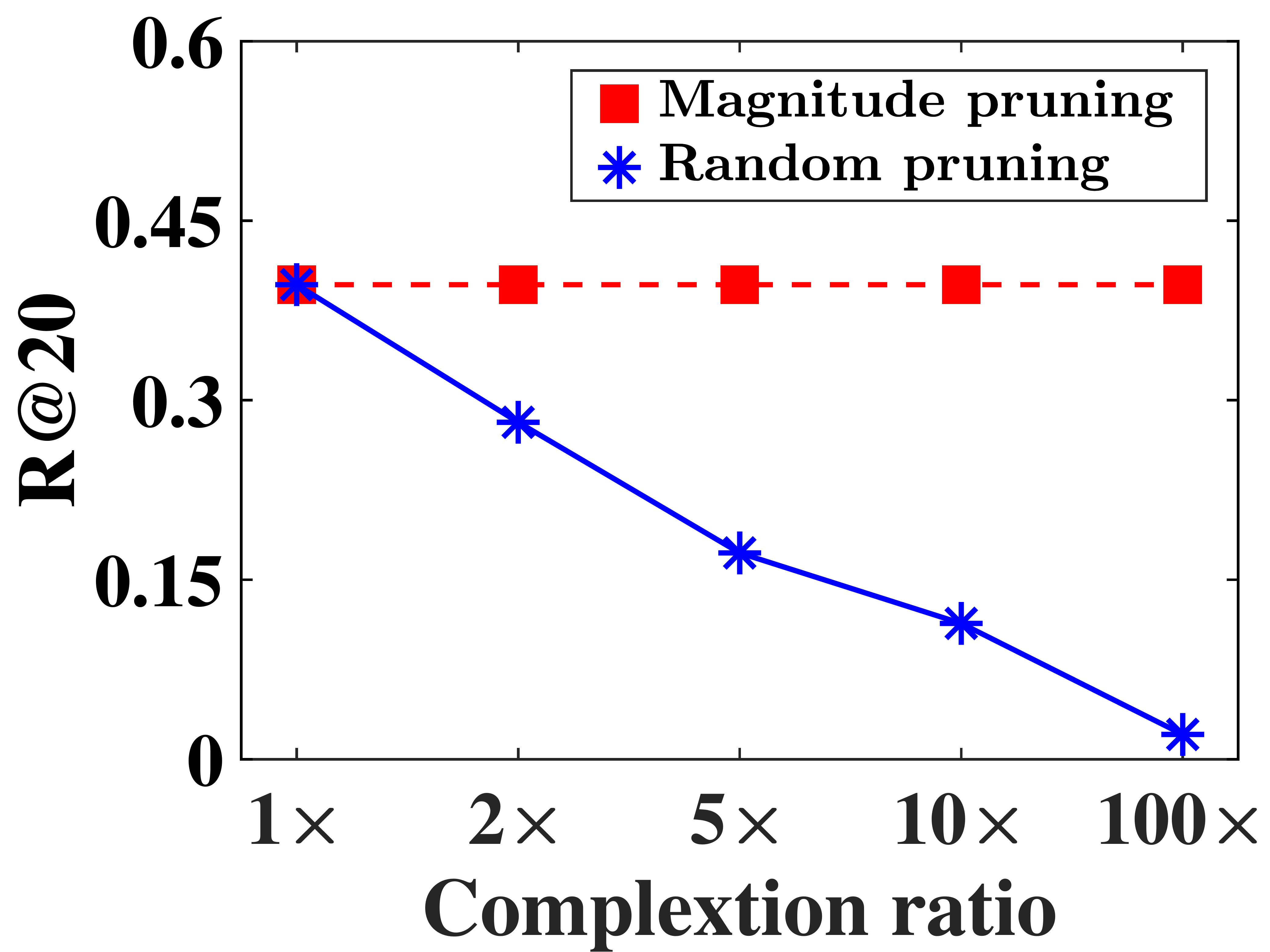} &
\includegraphics[width=0.20\textwidth]{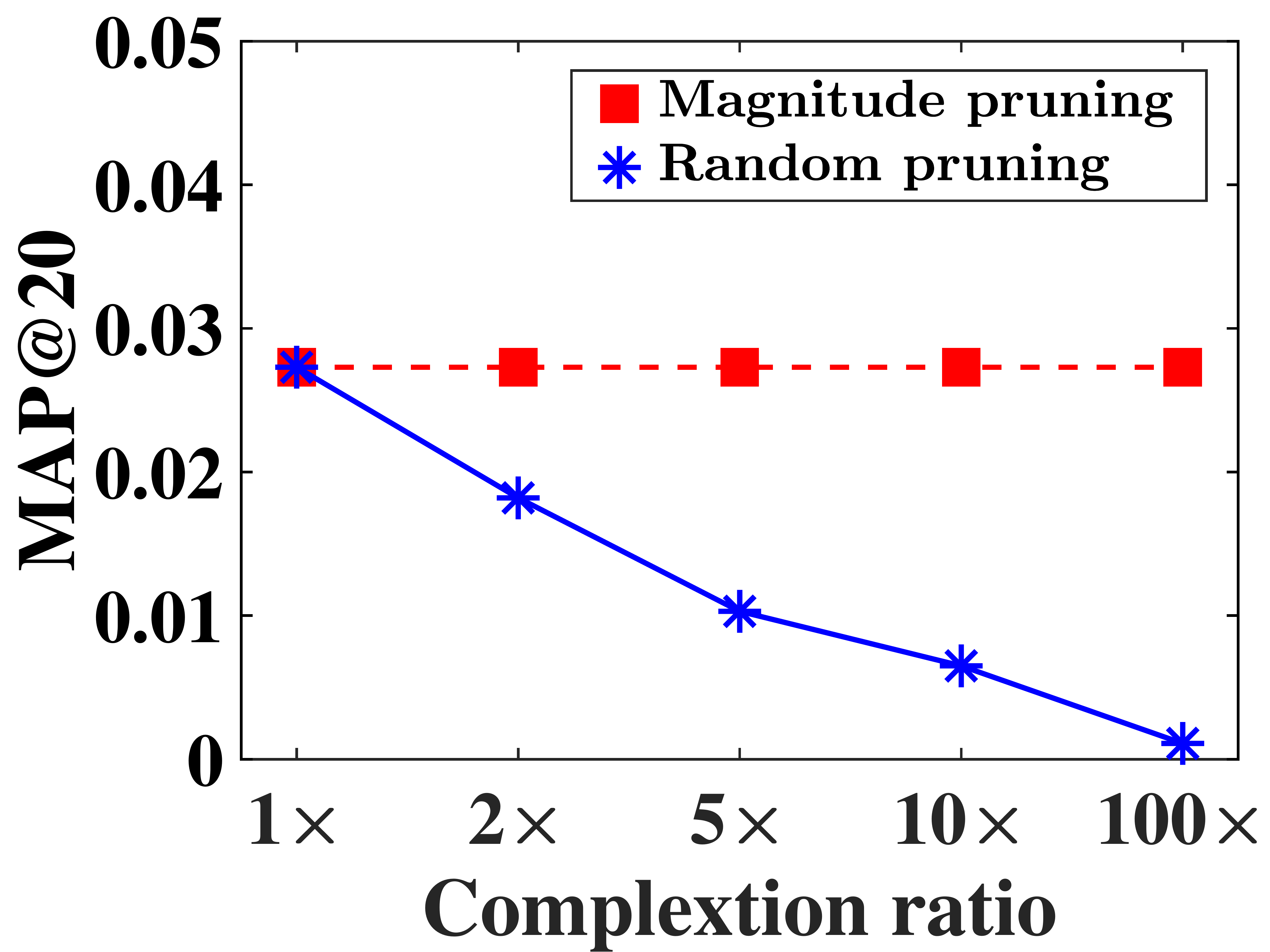}
\end{tabular} \\
(a) DIGI5

\begin{tabular}{cc}
\includegraphics[width=0.20\textwidth]{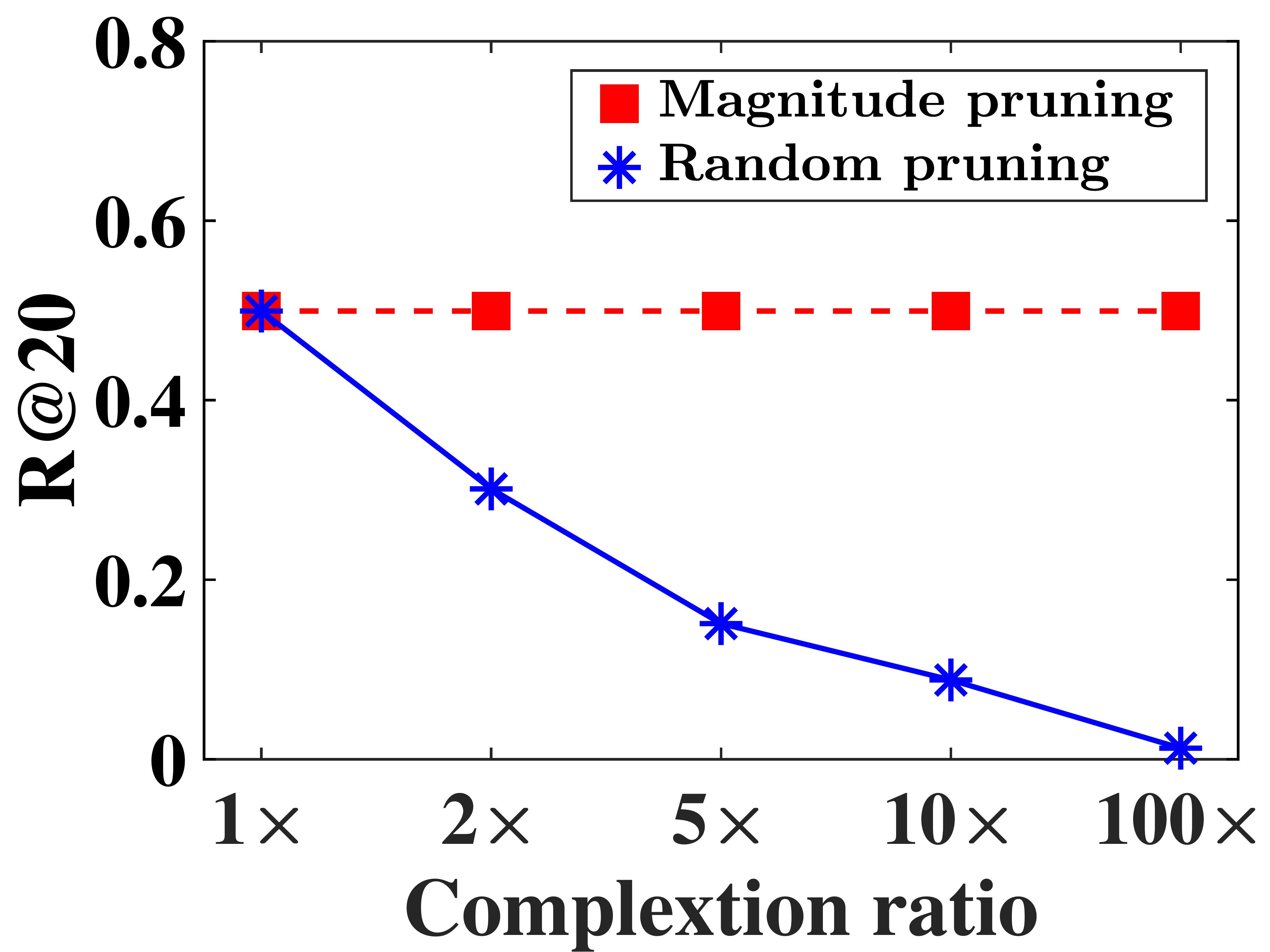} &
\includegraphics[width=0.20\textwidth]{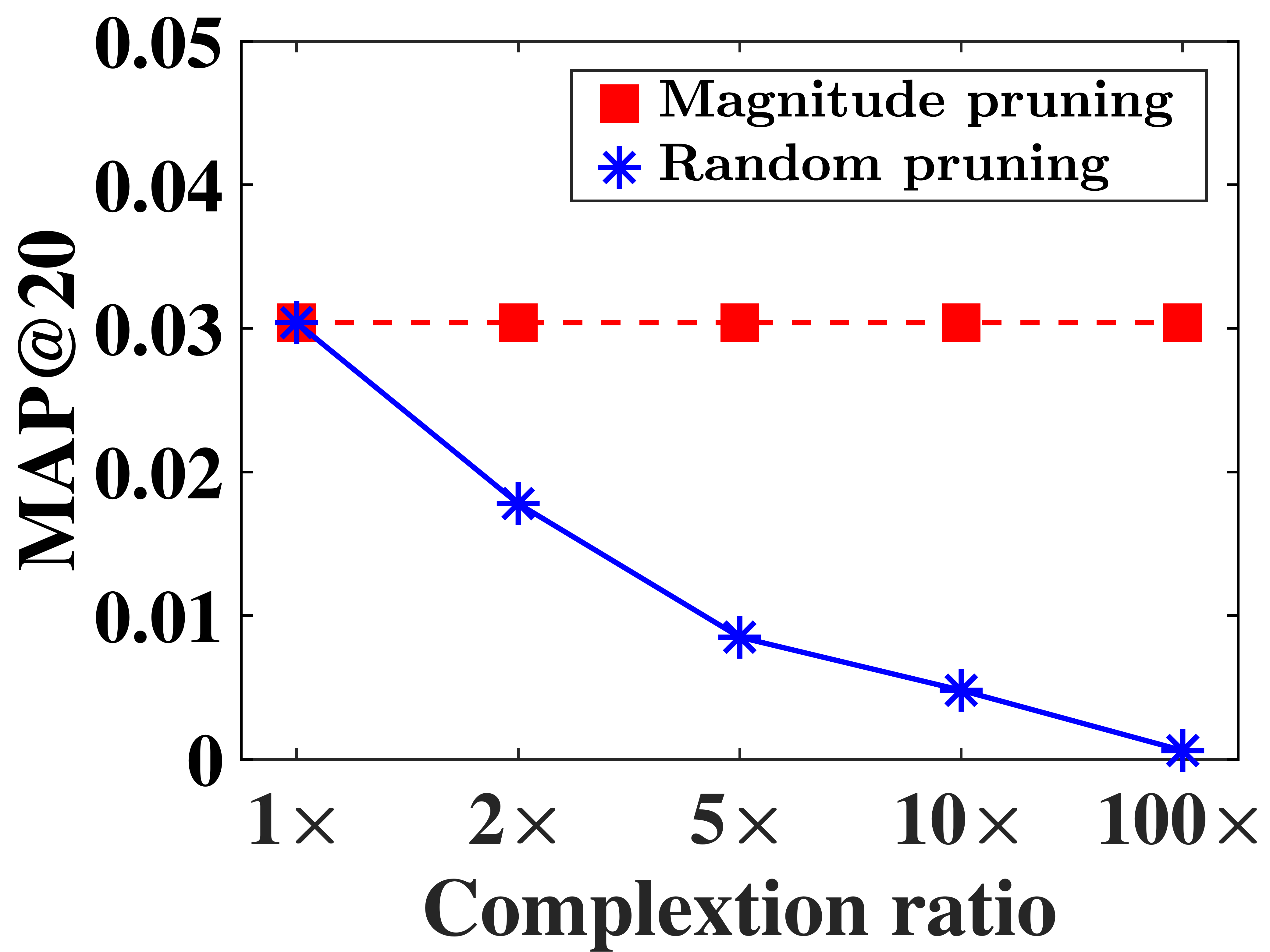}
\end{tabular} \\
(b) RR  \\


\vskip -0.1in
\caption{R@20 and MAP@20 of S-Walk over various compression ratios.
}
\label{fig:model_compression}
\end{figure}

%% file: Figures/Fig4_session_length_baseline.tex
\begin{figure}
\centering

\begin{tabular}{c}
\includegraphics[width=0.40\textwidth]{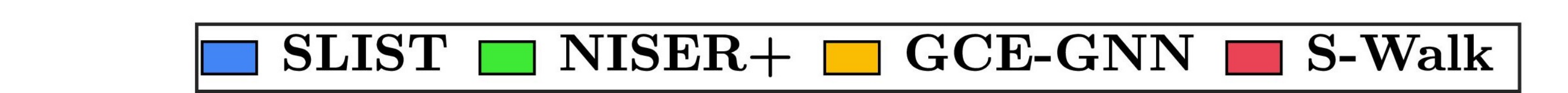}
\end{tabular} \\

\begin{tabular}{cc}
\includegraphics[width=0.20\textwidth]{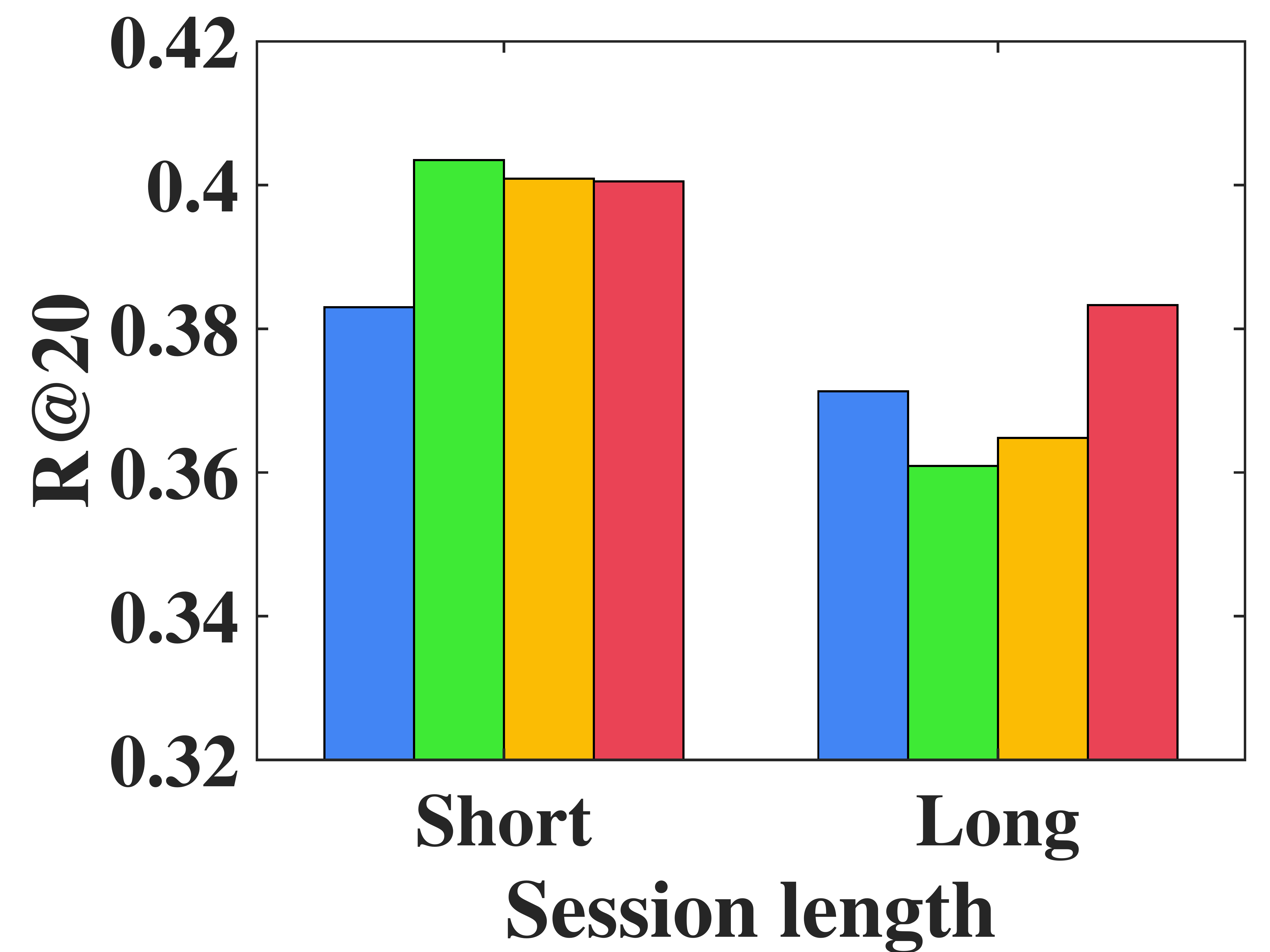} &
\includegraphics[width=0.20\textwidth]{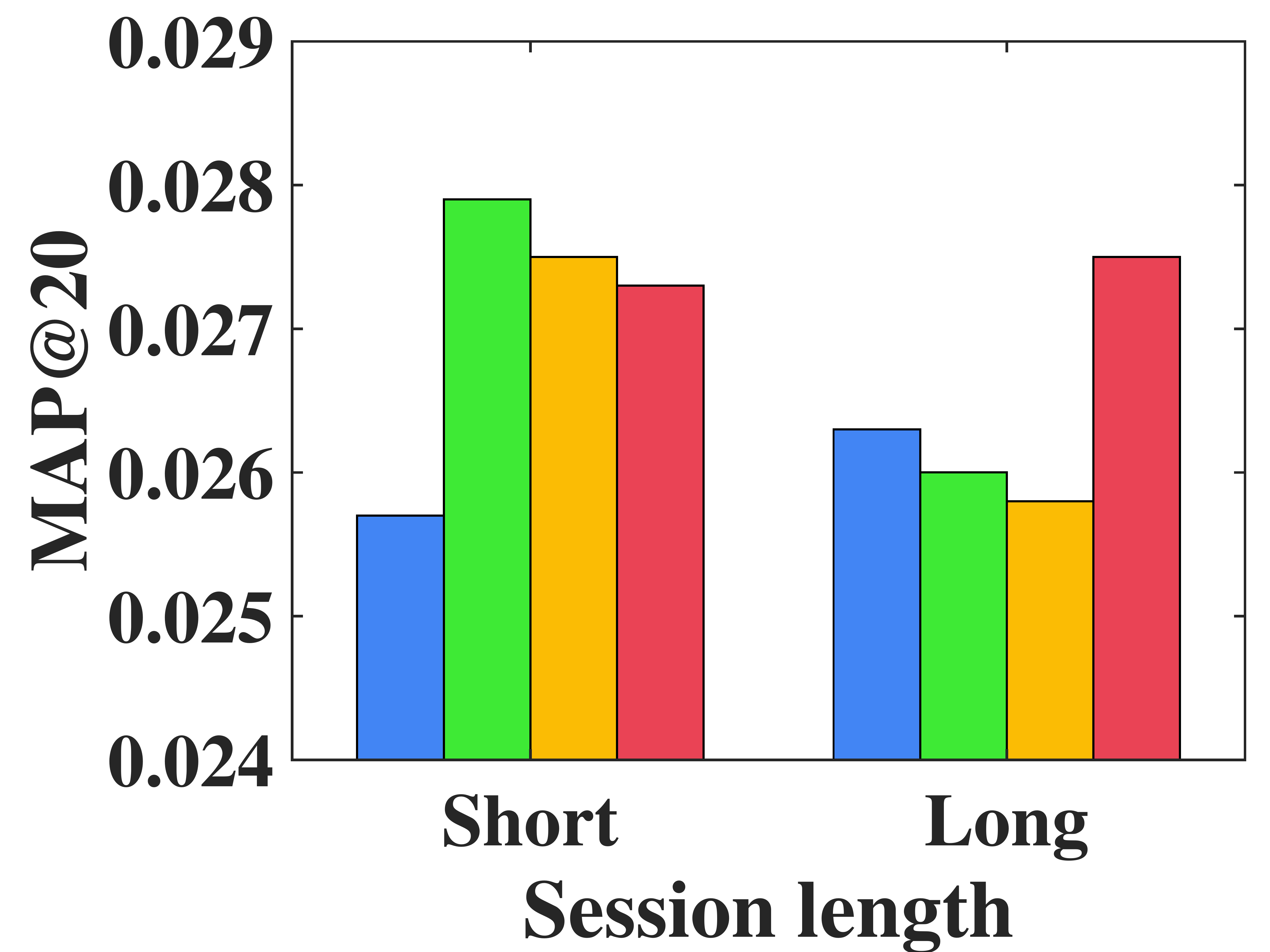}
\end{tabular} \\
(a) DIGI5  \\

\begin{tabular}{cc}
\includegraphics[width=0.20\textwidth]{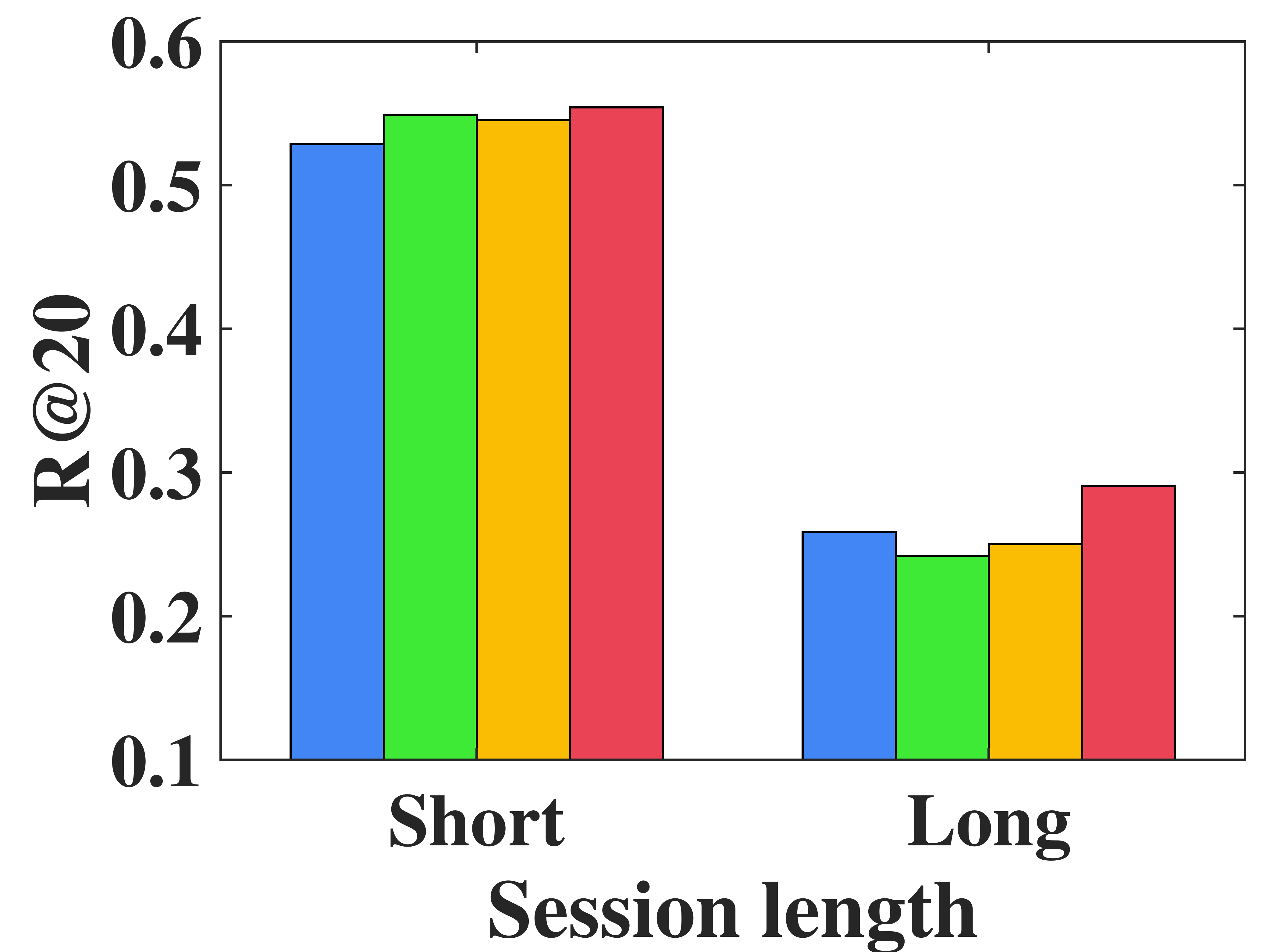} &
\includegraphics[width=0.20\textwidth]{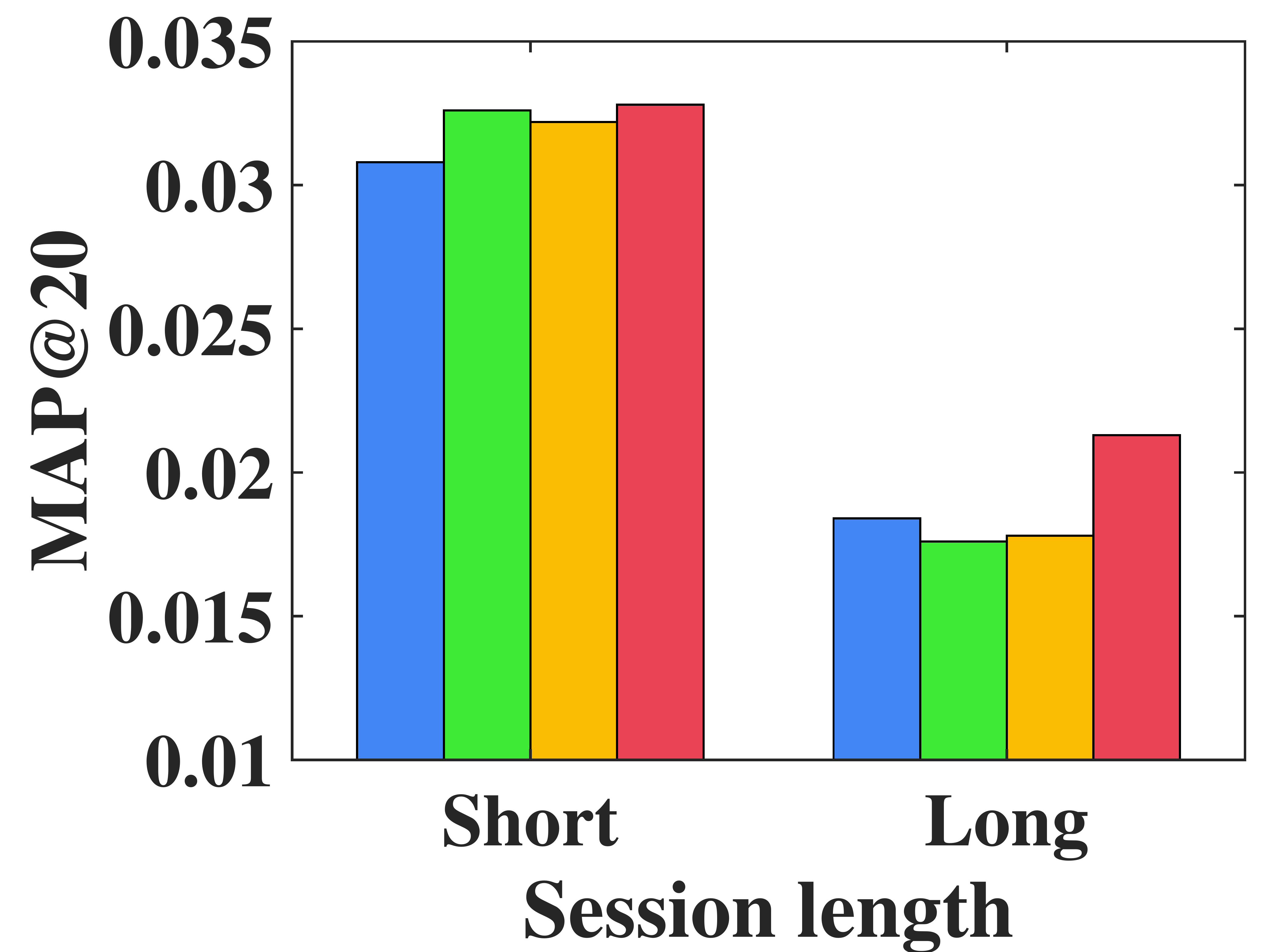}
\end{tabular} \\
(b) RR  \\

\vskip -0.1in
\caption{R@20 and MAP@20 of S-Walk and competitor models over different session lengths (Short $\le$ 5, Long > 5).}
\label{fig:length_baseline}
\vskip -0.15in
\end{figure}

%% file: Figures/Fig5_cutoff.tex
\begin{figure}
\centering

\begin{tabular}{c}
\includegraphics[width=0.40\textwidth]{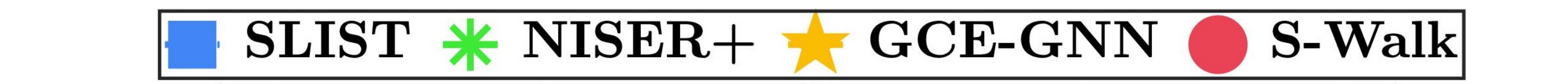}
\end{tabular} \\

\begin{tabular}{cc}
\includegraphics[width=0.20\textwidth]{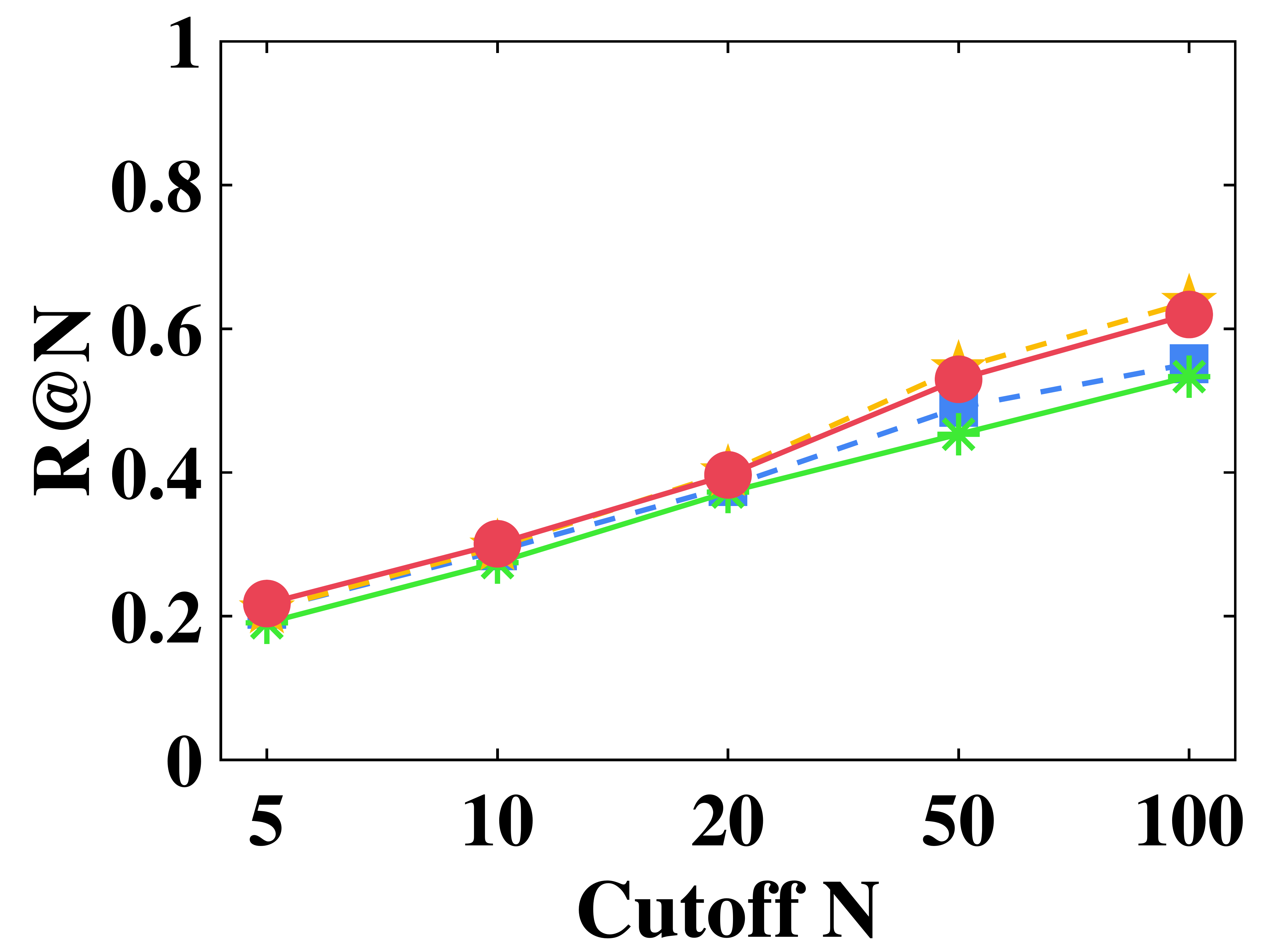} &
\includegraphics[width=0.20\textwidth]{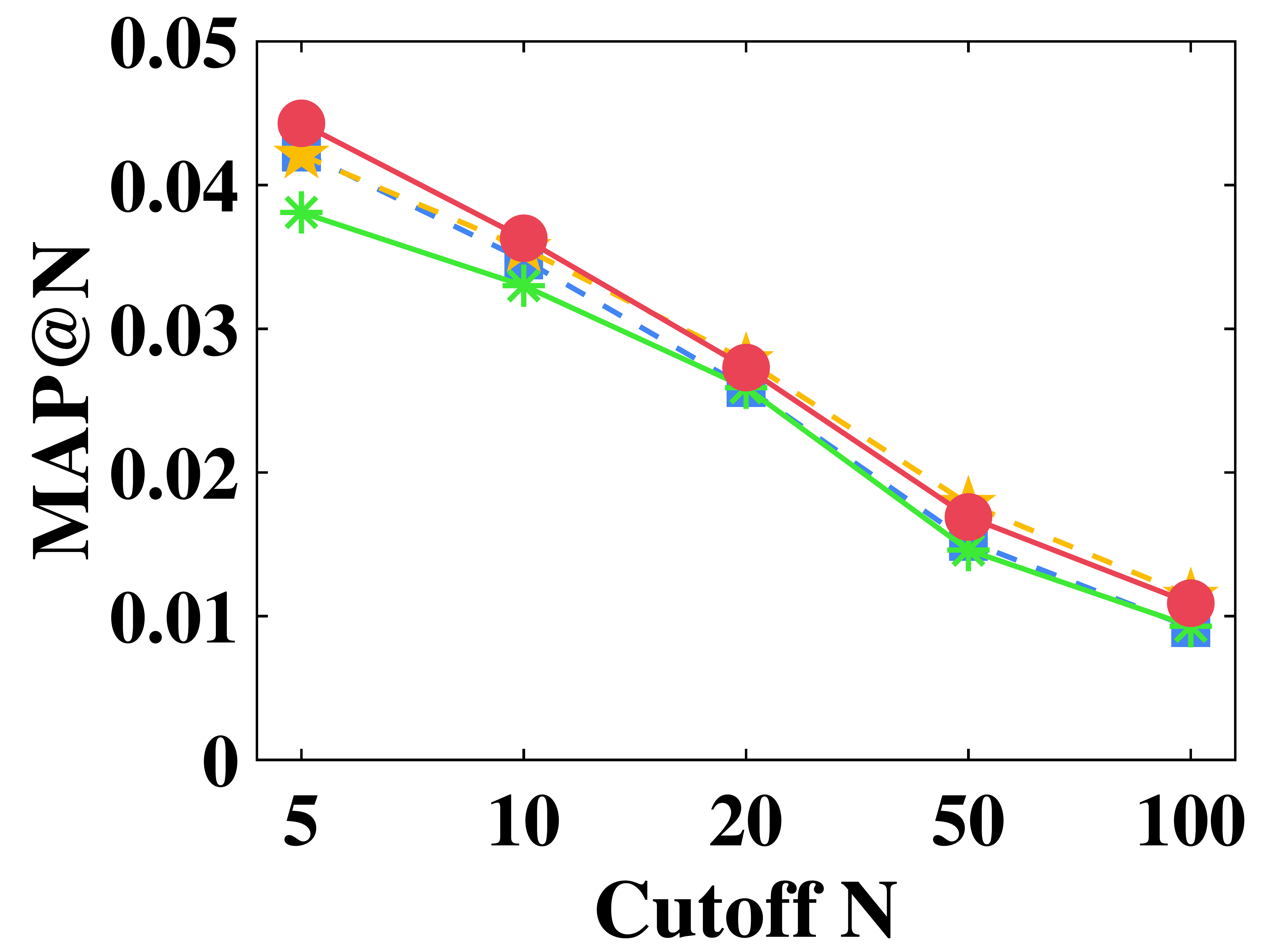}
\end{tabular} \\
(a) DIGI5  \\

\begin{tabular}{cc}
\includegraphics[width=0.20\textwidth]{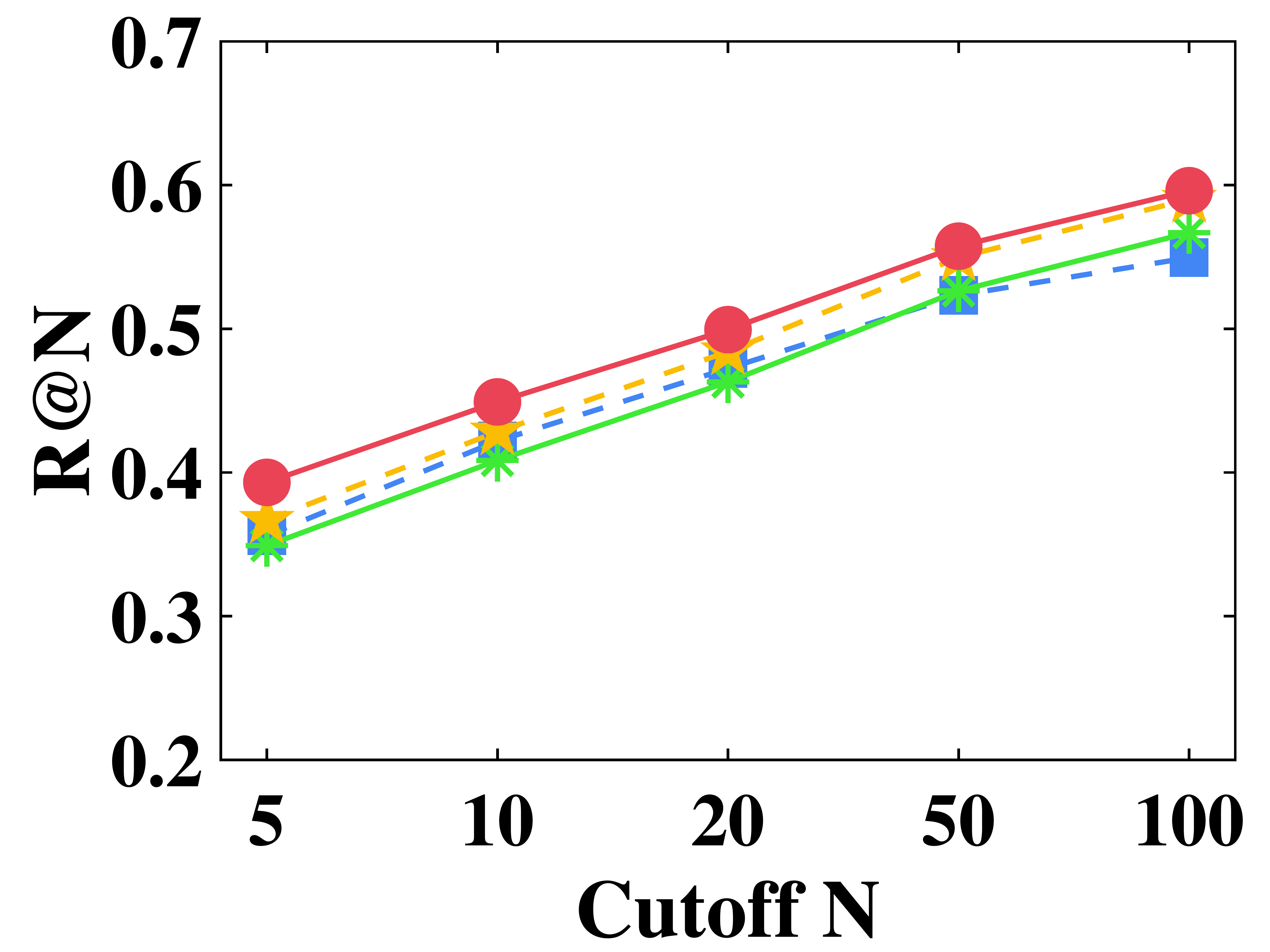} &
\includegraphics[width=0.20\textwidth]{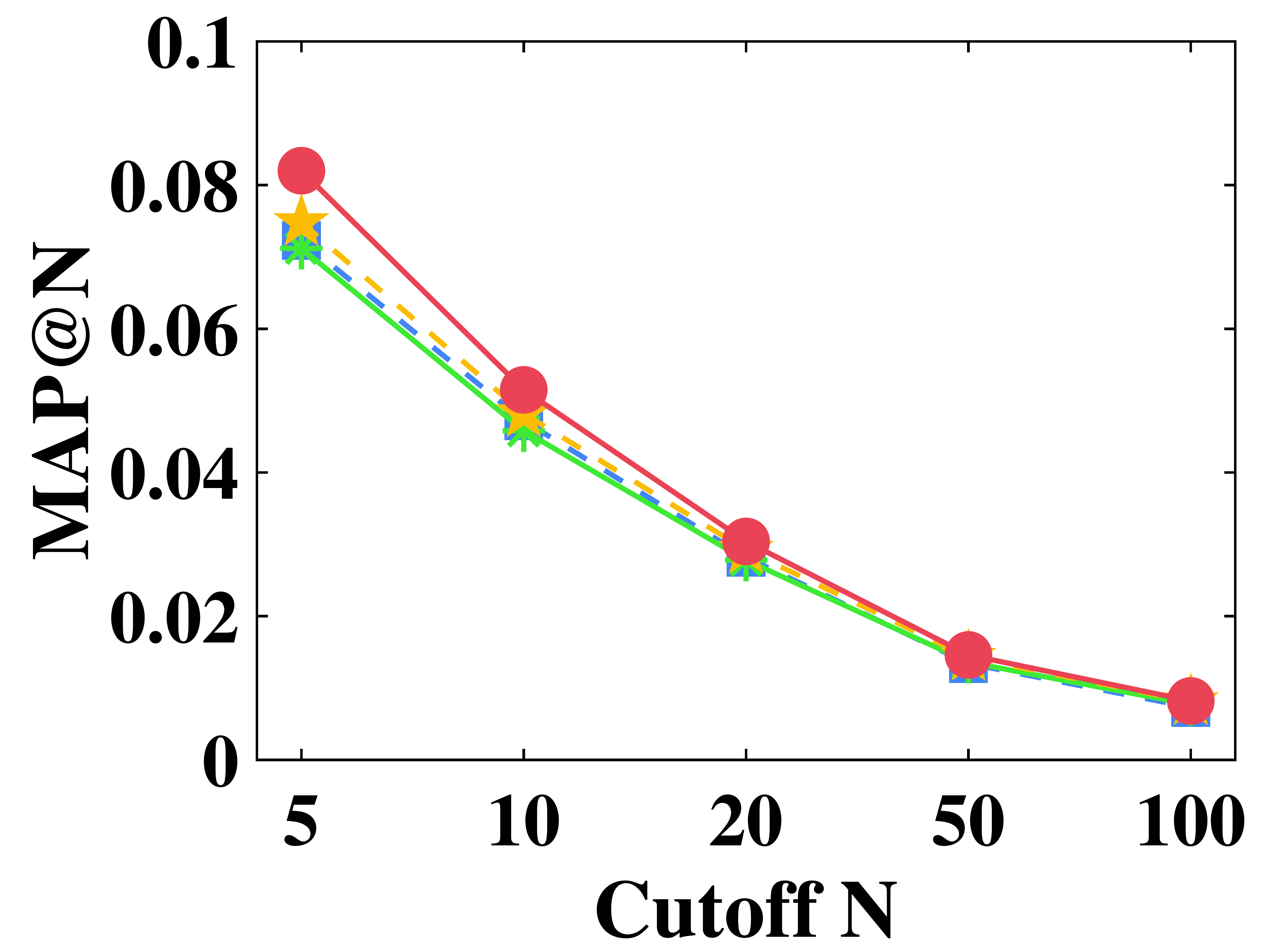}
\end{tabular} \\
(b) RR  \\

\vskip -0.1in
\caption{R@N and MAP@N of S-Walk and state-of-the-art models over the various cut-offs on DIGI5 and RR.}
\label{fig:cutoff}
\vskip -0.15in
\end{figure}

%% file: Tables/tab4_ablation.tex
\begin{table}[t] \small
\caption{R@20 and MAP@20 of S-Walk with various transition and teleportation models. SR~\cite{KamehkhoshJL17} and AR~\cite{AgrawalIS93} are simple Markov- and co-occurence-based models. $\mathbf{I}$ denotes the identity matrix. $tran$ and $tele$ denote the transition and teleportation model, respectively.}
\vskip -0.1in
\label{tab:ablation}
\begin{tabular}{cc|cc|cc|cc}
\toprule
\multirow{2}{*}{$tran$}    & \multirow{2}{*}{$tele$} & \multicolumn{2}{c|}{YC-1/4}       & \multicolumn{2}{c|}{DIGI5}        & \multicolumn{2}{c}{RR}           \\
                      &                    & R              & MAP             & R               & MAP             & R               & MAP             \\ \midrule
\multirow{3}{*}{SR}   & $\mathbf{I}$                  & 0.5109          & 0.0394          & 0.3809          & 0.0260          & 0.4812          & 0.0291          \\
                      & AR                 & 0.4952          & 0.0378          & 0.3879          & 0.0266          & 0.4817          & 0.0291          \\
                      & Ours               & 0.5171          & 0.0400          & 0.3930          & 0.0270          & 0.4950          & 0.0301          \\ \hline
\multirow{3}{*}{Ours} & $\mathbf{I}$                  & 0.5175          & 0.0399          & 0.3808          & 0.0259          & 0.4826          & 0.0292          \\
                      & AR                 & 0.5009          & 0.0383          & 0.3899          & 0.0268          & 0.4856          & 0.0293          \\
                      & Ours               & \textbf{0.5205} & \textbf{0.0403} & \textbf{0.3936} & \textbf{0.0271} & \textbf{0.4979} & \textbf{0.0303} \\ \bottomrule     
\end{tabular}
\end{table}

%% file: Figures/Fig7_alpha_hyperparameter.tex
\begin{figure}[t]
\centering

\begin{tabular}{cc}
\includegraphics[width=0.22\textwidth]{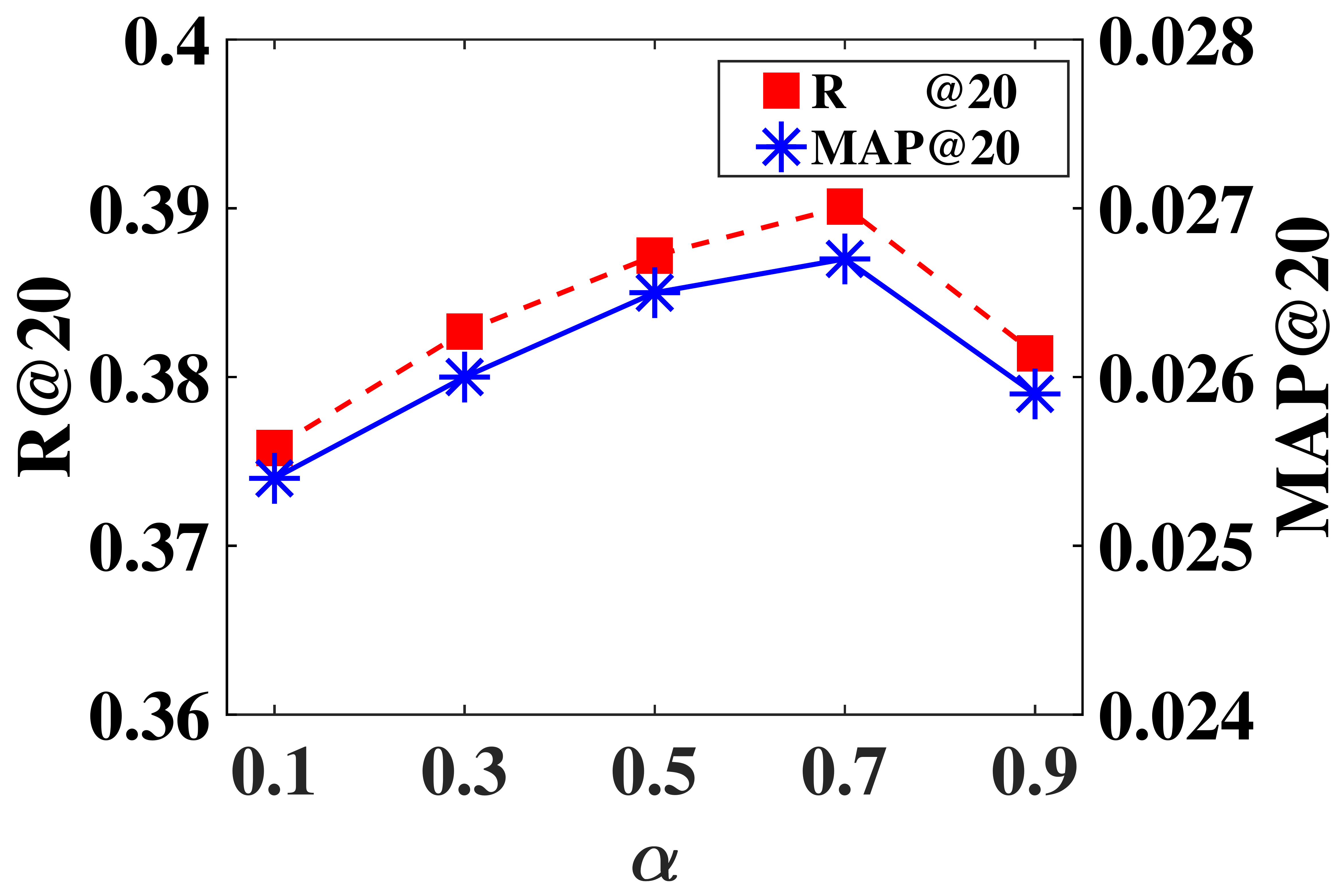} &
\includegraphics[width=0.22\textwidth]{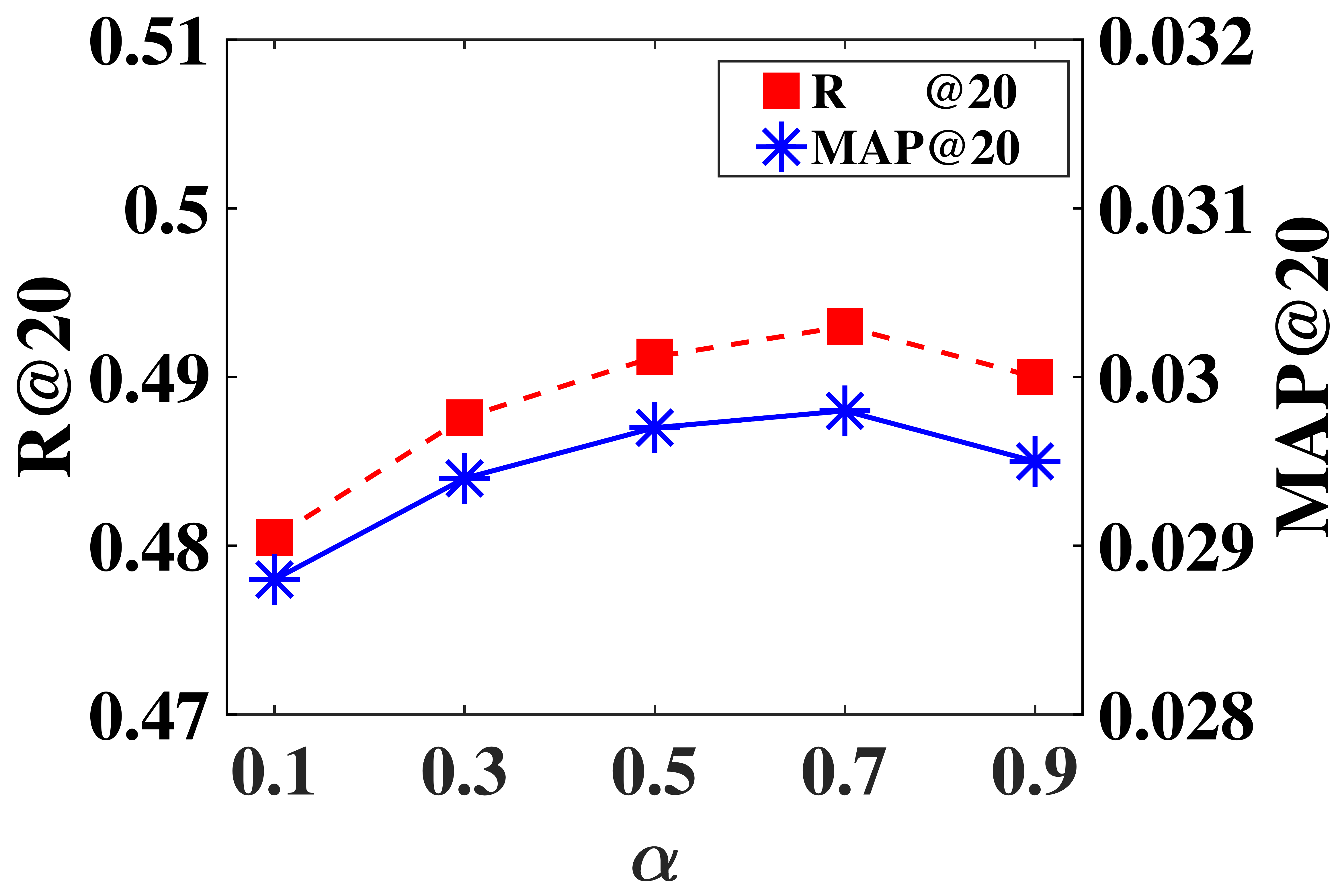} \\
(a) DIGI5 & (b) RR \\
\end{tabular}
\vskip -0.1in
\caption{R@20 and MAP@20 of S-Walk over varied damping factor $\alpha$.}
\label{fig:combine_alpha}
\vskip -0.1in
\end{figure}

%% file: sec-relatedwork.tex
\section{Related Work}\label{sec:relatedwork}

\vspace{1mm}
\noindent
\textbf{Random walk-based models}. With the success of PageRank~\cite{AmyNCarlD06pagerank}, the idea of the random walk has been widely adopted to address the data sparsity problem in recommender systems~\cite{YildirimK08}. TrustWalker~\cite{JamaliE09} addressed cold-start user/item problems using trust information between users. Starting from a target user vertex, \cite{CooperLRS14,ChristoffelPNB15} estimated the proximity score using the transition probabilities after short random walks on the user-item bipartite network. Besides, random-walk-based models have been successfully deployed in the large-scale industrial systems~\cite{EksombatchaiJLL18}. Recently, RecWalk~\cite{NikolakopoulosK19,NikolakopoulosK20} leveraged spectral properties of nearly decoupled Markov chains and combined the item model with random walks.
\input{Figures/Fig8_Swalk_step}

\vspace{1mm}
\noindent
\textbf{Session-based recommendation (SR)}. SR models are categorized into three groups, \ie, \emph{Markov chain models}, \emph{neighborhood-based models}, and \emph{DNN-based models}. For more details, please refer to recent survey papers~\cite{BonninJ14,JannachLL17}. Firstly, Markov chains (MC) are useful for modeling consecutive item dependency. FPMC~\cite{RendleFS10} proposed the tensor factorization using MF, and FOSSIL~\cite{HeM16} combined FISM~\cite{KabburNK13} with factorized MC. SR~\cite{KamehkhoshJL17} proposed an improved MC model by combining association rules. Although they are effective for addressing the short-term item dependency, it does not utilize various patterns among items. Secondly, \citet{JannachL17} adopted the K-nearest neighbor (KNN) for SR, and STAN~\cite{GargGMVS19} improved SKNN using various weight schemes to further reflect item dependency. Recently,~\cite{LudewigJ18,LudewigMLJ19a,LudewigMLJ19b} reported that the KNN-based models have shown competitive performance on various datasets. However, they are generally limited in representing high-order dependency among items. Lastly, various DNN-based models have been used for SR. GRU4Rec~\cite{HidasiKBT15, HidasiK18} employed gated recurrent units (GRU) for SR. Later, NARM~\cite{LiRCRLM17} and STAMP~\cite{LiuZMZ18} utilized attention mechanisms to distinguish short- and long-term item dependency. To further analyze complex item transitions, SR-GNN~\cite{WuT0WXT19} recently exploited gated graph neural networks (GGNN). However, SR-GNN~\cite{WuT0WXT19} is often vulnerable to overfitting~\cite{GuptaGMVS19} owing to extreme data sparsity.


%% file: Figures/Fig8_Swalk_step.tex
\begin{figure}[t]
\centering

\begin{tabular}{cc}
\includegraphics[width=0.20\textwidth]{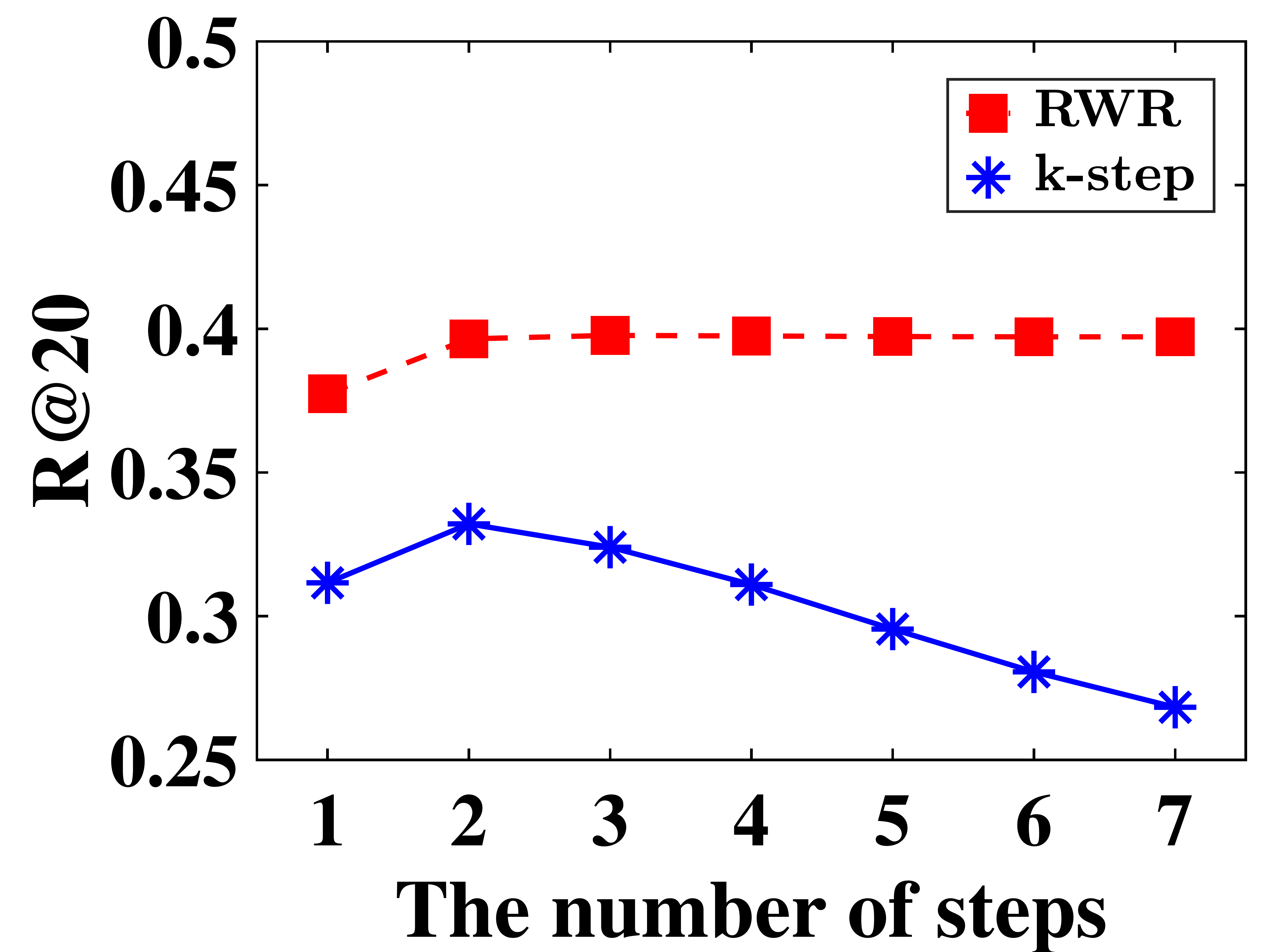} &
\includegraphics[width=0.20\textwidth]{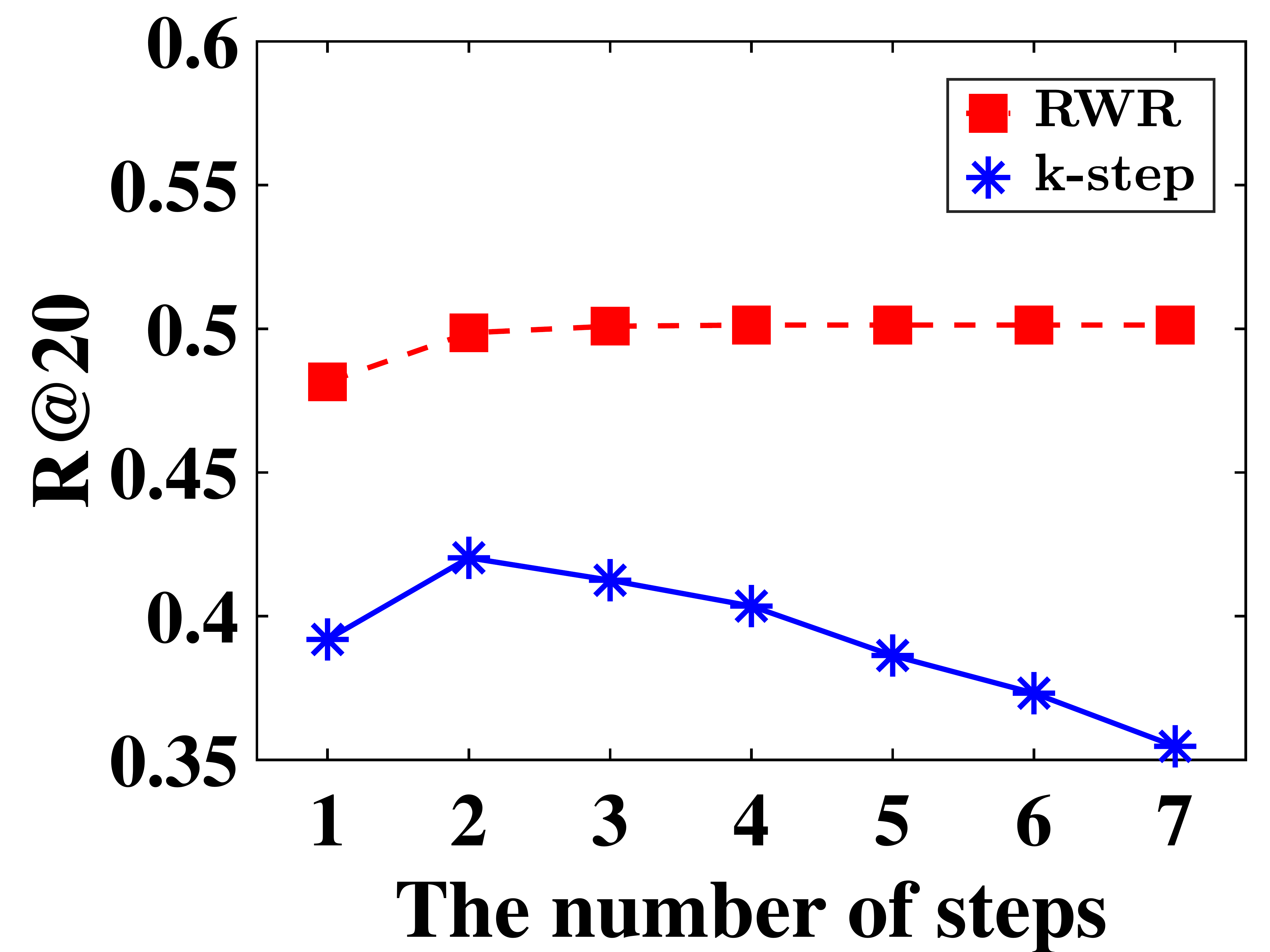} \\
(a) DIGI5 & (b) RR \\
\end{tabular}

\vskip -0.1in
\caption{R@20 of S-Walk using random walk with restart (RWR) and a k-step landing probability.}
\label{fig:steps}
\vskip -0.2in
\end{figure}

%% file: sec-conclusion.tex
\section{Conclusion}\label{sec:conclusion}

In this work, we propose a session-based recommendation using item random walks, \emph{S-Walk}. To complement the drawback of existing models, we utilize the random walk with restart to fully capture intra- and inter-correlations of sessions. We incorporate efficient linear item models, \ie, the transition model and the teleportation model, into the item random walks process. Our extensive evaluation shows that S-Walk achieves comparable or state-of-the-art accuracies, high scalability, and fast inference speed over various benchmark datasets.

\section*{Acknowledgment}

This work was supported by the National Research Foundation of Korea (NRF) (NRF-2018R1A5A1060031, NRF-2021R1F1A1063843, and NRF-2021H1D3A2A03038607). Also, this work was supported by Institute of Information \& communications Technology Planning \& evaluation (IITP) grant funded by the Korea government (MSIT) (No. 2019-0-00421, AI Graduate School Support Program).

%% file: sec-appendix.tex
\newpage
\appendix

\section{Closed-form Solutions}\label{sec:app_solution}

We provide a detailed derivation of the closed-form solutions for the item transition model and the item teleportation model in Section~\ref{sec:transtion_matrix} and \ref{sec:teleporation_matrix}. Based on the closed-form solutions, we can further apply for row-level weights to reflect the \emph{timeliness} of sessions, \ie, the more important, the more recent sessions.

\subsection{Linear Item Transition Model}

Given the input matrix $\mathbf{Y} \in \mathbb{R}^{m \times n}$ and the output matrix $\mathbf{Z} \in \mathbb{R}^{m \times n}$, the objective function is expressed as
\begin{equation}
  \label{eq:slit_loss4}
  \argmin_{\mathbf{B}^\text{tran}} \ \mathcal{L}(\mathbf{B}^\text{tran}) = \left\|( \mathbf{Z} - \mathbf{Y} \cdot \mathbf{B}^\text{tran} )\right\|_F^2 + \lambda \|\mathbf{B}^\text{tran}\|_F^2.
\end{equation}

\noindent
The first-order derivative of Eq.~\eqref{eq:slit_loss4} over $\mathbf{B}^{tran}$  is then given by
\begin{equation}
\label{eq:B_hat_of_SLIT}
\begin{aligned}
   \frac{1}{2} \cdot \frac{\partial \mathcal{L}}{\partial \mathbf{B}^\text{tran} }  &  \ =  \ \ (-\mathbf{Y}^{\top}){(\mathbf{Z} - \mathbf{Y} \cdot \mathbf{B}^\text{tran})} + \lambda \mathbf{B}^\text{tran}, \\
    & \ =  \ \  (\mathbf{Y}^{\top} \mathbf{Y} + \lambda\mathbf{I})\cdot\mathbf{B}^\text{tran} - \mathbf{Y}^{\top} \mathbf{Z}.
\end{aligned}
\end{equation}

Solving for $\mathbf{B}^\text{tran}$ so that Eq.~\eqref{eq:B_hat_of_SLIT} becomes $0$ gives the closed-form solution of Eq.~\eqref{eq:slit_loss4}:
\begin{equation}
  \label{eq:slit_solution4}
  \hat{\mathbf{B}}^\text{tran} \ = \ \hat{\mathbf{P}}' \cdot (\mathbf{Y}^{\top} \mathbf{Z}),
\end{equation}
where $\hat{{\mathbf{P}}}' = \left( \mathbf{Y}^{\top} \mathbf{Y} + \lambda \mathbf{I} \right)^{-1}$. 

\subsection{Linear Item Teleportation Model}

\noindent
To solve the constrained optimization problem, we define a new objective function $\mathcal{L}(\mathbf{B}^\text{tele}, \mathbf{\mu})$ by applying a Lagrangian multiplier and a KKT condition:
\begin{equation}
  \label{eq:slis_loss2}
  \begin{aligned}
    \mathcal{L}(\mathbf{B}^\text{tele}) = \ & \| ( \mathbf{X} - \mathbf{X} \cdot \mathbf{B}^\text{tele} )\|_F^2 + \lambda \|\mathbf{B}^\text{tele}\|_F^2 \\ & \ \ \text{s.t.} \ \ \texttt{diag}(\mathbf{B}^\text{tele}) \le \xi \\
  \end{aligned}
\end{equation}

\begin{equation}
  \label{eq:SLIS_kkt_loss}
  \begin{aligned}
    \mathcal{L}(\mathbf{B}^\text{tele}, \mu) = \ & \| ( \mathbf{X} - \mathbf{X} \cdot \mathbf{B}^\text{tele} )\|_F^2 + \lambda \|\mathbf{B}^\text{tele}\|_F^2 \\
    & + 2\mathbf{\mu}^{\top}\texttt{diag}(\mathbf{B}^\text{tele} - \xi\mathbf{I}),
  \end{aligned}
\end{equation}
where $\mathbf{\mu} \in \mathbb{R}^{n}$ is the KKT multiplier that satisfies $\forall \text{i}, \mathbf{\mu}_{i} \geq 0$. Then, we differentiate $\mathcal{L}(\mathbf{B}^\text{tele}, \mathbf{\mu})$ with respect to $\mathbf{B}^\text{tele}$ to minimize Eq.~\eqref{eq:SLIS_kkt_loss}:
\begin{align}
    \frac{1}{2}\frac{\partial \mathcal{L} (\mathbf{B}^\text{tele},\mu) }{\partial \mathbf{B}^\text{tele} }  &  \ = (-\mathbf{X}^{\top})  (\mathbf{X} - \mathbf{X} \cdot \mathbf{B}^\text{tele}) + \lambda\mathbf{B}^\text{tele} + \texttt{diagMat}(\mathbf{\mu}) \nonumber
  \label{eq:SLIS_kkt_loss_diff} \\
    & \ =    - \mathbf{X}^{\top}  \mathbf{X} + \mathbf{X}^{\top}  \mathbf{X} \cdot \mathbf{B}^\text{tele} + \lambda\mathbf{B}^\text{tele} + \texttt{diagMat}(\mathbf{\mu})  \nonumber \\
    & \ =   (\mathbf{X}^{\top}  \mathbf{X} + \lambda\mathbf{I})\mathbf{B}^\text{tele} - \mathbf{X}^{\top}  \mathbf{X} +\texttt{diagMat}(\mathbf{\mu}). 
\end{align}

Setting this to 0 and solving by $\mathbf{B}^\text{tele}$ gives the optimal $\hat{\mathbf{B}}^\text{tele}$ as
\begin{align}
    \hat{\mathbf{B}}^\text{tele} & = (\mathbf{X}^{\top}  \mathbf{X} + \lambda\mathbf{I})^{-1} \cdot  [\mathbf{X}^{\top}  \mathbf{X}+ \lambda\mathbf{I} - \lambda\mathbf{I} -\texttt{diagMat}(\mathbf{\mu})] \nonumber\\
    & = \hat{\mathbf{P}}[\hat{\mathbf{P}}^{-1} - \lambda\mathbf{I} - \texttt{diagMat}(\mathbf{\mu}) ] \nonumber\\
    & = \mathbf{I} - \hat{\mathbf{P}}[\lambda\mathbf{I} + \texttt{diagMat}(\mathbf{\mu}) ] \nonumber \\ 
    & = \mathbf{I} - \lambda \hat{\mathbf{P}} - \hat{\mathbf{P}}\cdot \texttt{diagMat}(\mathbf{\mu}), 
    \label{eq:SLIS_optimal_B_derivation}
\end{align}
where $\hat{\mathbf{P}} = \left( \mathbf{X}^{\top} \mathbf{X} + \lambda\mathbf{I} \right)^{-1}$.

Besides, a KKT multiplier $\mathbf{\mu}_{j}$ is zero only if $\mathbf{B}^\text{tele}_{jj} \leq \xi$. Otherwise, $\mathbf{\mu}_{j}$ has a non-zero value. In this case, $\mathbf{\mu}_{i}$ regularizes the value of $\mathbf{B}^\text{tele}_{jj}$ as $\mathbf{B}^\text{tele}_{jj} = \xi$. For $\mathbf{B}^\text{tele}_{jj}$, we can develop the the following equation:
\begin{equation}
  \label{eq:regulate_by_mu}
  \mathbf{B}^\text{tele}_{jj} \ = \xi = \ 1 -  \lambda P_{jj} - P_{jj}\mathbf{\mu}_{j}.
\end{equation}

Finally, $\mathbf{\mu}_{j}$ can be expressed as follows:

\begin{equation}
   \mu_{j} = \frac{(1 - \lambda P_{jj} - \xi )}{P_{jj}}
     = \frac{(1 - \xi)}{P_{jj}} - \lambda.
\end{equation}

Substituting $\mu$ in Eq.~\eqref{eq:SLIS_optimal_B_derivation} and enforcing non-negative elements in $\hat{\mathbf{B}}^\text{tele}$ give $\hat{\mathbf{B}}^\text{tele}$:
\begin{equation}
  \label{eq:slis_solution4}
  \hat{\mathbf{B}}^\text{tele} = \mathbf{I} -  \hat{\mathbf{P}} \cdot  \texttt{diagMat}(\gamma),
\end{equation}
\begin{equation}
    \label{eq:case_of_gamma_App}
    \gamma_{j} =
    \begin{cases}
        \ \ \ \lambda & \text{if} \ \ \ 1 - \lambda P_{jj} \le \xi \\
        \ \ \ \frac{1-\xi}{P_{jj}} & \text{otherwise}. 
    \end{cases}
\end{equation}
Here, $\mathbf{\gamma}$ is a vector defined by $\mathbf{\gamma} = \mathbf{\mu} + \lambda \cdot \mathbf{1}$.

\section{Training of S-Walk}\label{sec:algorithm}

Algorithm \ref{alg:training} describes how the final item-item matrix $\mathbf{M}$ is trained using the transition matrix $\mathbf{R}$ and the teleportation matrix $\mathbf{T}$. First, we define $\mathbf{M}_{(0)}$ as the identity matrix. Then, we repeatedly update $\mathbf{M}_{(k)}$ using $\mathbf{R}$ and $\mathbf{T}$. For example, $\mathbf{M}_{(1)}$ can be thought as $\mathbf{R}$ with a probability of $\alpha$ and $\mathbf{T}$ with a probability of $(1-\alpha)$. 
After $\mathbf{M}_{(k)}$ converges, we use $\mathbf{M}_{(k)}$ as the item-item matrix for inference.

\input{algorithm_training}

\section{Theoretical Analysis}\label{sec:analysis}

\noindent
\textbf{Property of the item transition matrix}. The item transition matrix $\mathbf{R}$ can be viewed as a generalized Markov model. When representing partial representations, the importance of items can be different. As $\delta_{\text{pos}}$ in Eq.~\eqref{eq:weight_item} is small, it can be close to a simple Markov model~\cite{KamehkhoshJL17}, capturing consecutive transitions between items. We can also consider a broader range to represent the transition between items. While the simple Markov model is mostly based on the frequency of consecutive items, our linear model can estimate the sequential correlation between items. As a result, it can outperform existing Markov models~\cite{RendleFS10, KabburNK13, HeM16, KamehkhoshJL17}.

\vspace{1mm}
\noindent
\textbf{Property of the item teleportation matrix}. As discussed in~\cite{Steck19b}, assume that a training matrix $\mathbf{X}$ is $|\mathcal{S}|$ samples with $|\mathcal{I}|$ random variables, \ie, $x \sim \mathcal{N}(0, \mathbf{\Sigma})$ following a Gaussian distribution with zero mean and the covariance matrix $\mathbf{\Sigma} \in \mathbb{R}^{n \times n}$. Here, the estimate of the covariance matrix is $\hat{\mathbf{\Sigma}} = \mathbf{X}^{\top} \mathbf{X} / |\mathcal{S}|$, and $\hat{\mathbf{P}} = \hat{\mathbf{\Sigma}}^{-1}$ is the estimate of the precision (or concentration) matrix. By solving the closed-form equation in Eq.~\eqref{eq:tele_solution}, the item teleportation matrix $\mathbf{T}$ can be interpreted as the precision matrix by estimating the co-occurrence matrix $\mathbf{X}^{\top} \mathbf{X} / |\mathcal{S}|$. In other words, the precision matrix can be viewed as the similarity matrix between items, which has been used in the recent neighborhood-based approach~\cite{VerstrepenG14, VolkovsY15}.

\vspace{1mm}
\noindent
\textbf{Model convergence}. The graph $\mathcal{G}_\mathbf{M}$ is defined by a combination of $\mathbf{R}$ and $\mathbf{T}$. The non-negative values in both matrices imply the positive correlations in covariance matrix $\mathbf{R}^{\top} \mathbf{T}$. The edges based on the correlations make graph $\mathcal{G}_\mathbf{M}$. Besides, $\mathbf{T}$ has a self-loop connection as in Eq.~\eqref{eq:tele_matrix}. Based on these structures, the transition matrix $\mathbf{R}$ is an ergodic Markov model, \ie, irreducible and aperiodic. That is, the landing probabilities of S-Walk converge to a limiting distribution. It is found that S-Walk converges by 3--5 steps, as observed in \ref{sec:exp_ablation}.

\section{Dataset Description}\label{sec:dataset}

Table~\ref{tab:statistics} summarizes the detailed statistics of all the benchmark datasets. These datasets are collected from e-commerce and music streaming services:
YooChoose\footnote{\url{https://www.kaggle.com/chadgostopp/recsys-challenge-2015}} (YC), Diginetica\footnote{\url{https://competitions.codalab.org/competitions/11161}} (DIGI), RetailRocket\footnote{\url{https://www.kaggle.com/retailrocket/ecommerce-dataset}} (RR), and NowPlaying\footnote{\url{https://drive.google.com/drive/folders/1ritDnO_Zc6DFEU6UND9C8VCisT0ETVp5}} (NOWP). For YC and DIGI, we use single-split datasets (\ie, YC-1/4 and DIGI1), following existing studies~\cite{HidasiK18, LiRCRLM17, LiuZMZ18, WuT0WXT19}. To evaluate on large-scale datasets, we further experiment on five-split datasets (\ie, YC, DIGI5, RR, NOWP), used in the recent empirical analysis~\cite{LudewigJ18,LudewigMLJ19a,LudewigMLJ19b}. For single-split datasets, we use the last day as the test set for YC-1/4 and the sessions of the last seven days as the test set for DIGI1, as done in~\cite{LudewigJ18, LudewigMLJ19a, LudewigMLJ19b}. For five-split datasets, we divide the datasets into five disjoint successive splits. For each split, the last $N$-days sessions were used for the test set. We choose $N$ for the test set as follows: one for YC, two for RR, five for NOWP, and seven for DIGI.

\input{Tables/tab1_dataset}

\section{Reproducibility}\label{sec:app_reproducibility}

\textbf{Implementation details}. For S-Walk, we tuned $\alpha$ and $\beta$  among \{0.1, 0.3, 0.5, 0.7, 0.9\}, $\delta_\text{pos}$ and $\delta_\text{inf}$ among \{0.125, 0.25, 0.5, 1, 2, 4, 8\} using the validation set selected from the training set for the same period as the testing set. For the baseline models, we used the best hyper-parameters reported in~\cite{LudewigJ18, LudewigMLJ19b}. We implemented the proposed model and STAN~\cite{GargGMVS19} using NumPy. We used public source code for SR-GNN\footnote{https://github.com/CRIPAC-DIG/SR-GNN}~\cite{WuT0WXT19} and implemented NISER+~\cite{GuptaGMVS19} using SR-GNN code. We used all source code for the other baseline models\footnote{https://github.com/rn5l/session-rec} in~\cite{LudewigMLJ19a}. For reproducibility, we conducted the reported results for all the baseline models, and actually verified whether the performance of baseline models could be reproduced with an error of 1--2\% or less in our implementation environments. We conducted all experiments on a desktop with 2 NVidia TITAN RTX, 256 GB memory, and 2 Intel Xeon Processor E5-2695 v4 (2.10 GHz, 45M cache).

\vspace{1mm}
\noindent
\textbf{Hyper-parameter setting}. Table~\ref{tab:hyperparameter} reports all hyper-parameters for S-Walk. Firstly, the hyper-parameters of the left column are the factor that controls the involvement of random walk models. $\alpha$ is a damping factor that determines the proportion of the transition matrix $\mathbf{S}$ and teleportation matrix $\mathbf{T}$. $\beta$ controls the probability that the surfer jumps to the item rather than to itself. Lastly, the hyper-parameters of the right column are model hyper-parameters used for the transition model and the teleportation model.


\input{Tables/tab5_optimal_param}

\section{Additional Experimental Results}\label{sec:app_additionalresults}

\input{Figures/Fig6_train_size}

We also evaluate the accuracy of S-Walk and the competing models by varying the size of the entire session from 10--100\%. As shown in Figure~\ref{fig:train_size}, S-Walk consistently achieves better accuracy than other baselines. Even though all models degrade performance with a smaller training set, S-Walk suffers the least from sparser data. Notably, DNN-based models show significant degradation in accuracy with smaller training sets owing to their complex structures and enormous parameters. (For the other datasets, we still observe similar tendencies.) Based on this observation, S-Walk is still more effective than the baseline models, even on small-scale datasets.





%% file: algorithm_training.tex
 \vspace{-3mm}
\SetKwInOut{Parameter}{Parameters}
\SetEndCharOfAlgoLine{}
\begin{algorithm}[h]
  \KwIn{Transition matrix $\mathbf{R}$, teleportation matrix $\mathbf{T}$.}
  \KwOut{S-Walk model $\mathbf{M}$.}
  $\mathbf{M}_{(0)} \leftarrow \mathbf{I}$ \\
  $k = 0$ \\
  
  \Repeat{$ \| \mathbf{M}_{(k)} - \mathbf{M}_{(k-1)} \|_{1} \leq \epsilon $}
  {
  $k \leftarrow k+1$\\
  $\mathbf{M}_{(k)} \leftarrow \alpha\mathbf{M}_{(k-1)}\mathbf{R} + (1-\alpha)\mathbf{T} $ \\
  }

  $\mathbf{M} \leftarrow \mathbf{M}_{(k)}$ \\
\caption{Training procedure for S-Walk}\label{alg:training}
\end{algorithm}
\vspace{-3mm}


%% file: Tables/tab1_dataset.tex
\begin{table}
\small
\caption{Statistics of the benchmark datasets. \#Actions indicates the number of entire user-item interactions.}
\vspace{-2mm}
\label{tab:statistics}
\begin{tabular}{ll|rrrrr}
\toprule
Split & Dataset & \#Actions & \#Sessions & \#Items & \#Actions & \#Items \\
& & & & & / Sess. & / Sess.\\
\midrule
1-split
& YC-1/4  & 7,909,307 & 1,939,891  & 30,638  & 4.08          & 3.28        \\
& DIGI1  & 916,370   & 188,807    & 43,105  & 4.85          & 4.08        \\
\midrule
5-split
& YC5      & 5,426,961 & 1,375,128  & 28,582  & 3.95          & 3.17        \\
& DIGI5  & 203,488   & 41,755     & 32,137  & 4.86          & 4.08        \\
& RR      & 212,182   & 59,962     & 31,968  & 3.54          & 2.56        \\
& NOWP    & 271,177   & 27,005     & 75,169  & 10.04         & 9.38        \\
\bottomrule
\end{tabular}
\vspace{-3mm}
\end{table}

%% file: Tables/tab5_optimal_param.tex

\begin{table}
\caption{Hyper-parameter settings of S-Walk. $\lambda$ is a L2-weight decay, $\alpha$ is a damping factor and $\beta$ is the self-loop factor. $\delta_\text{pos}$ is a weight decay by item position and $\delta_\text{inf}$ is a weight decay by inference item position.}
\vspace{-2mm}
\label{tab:hyperparameter}
\begin{tabular}{c|cc|ccc}
\toprule
 Dataset & $\alpha$ & $\beta$ & $\lambda$ & $\delta_\text{pos}$ & $\delta_\text{inf}$ \\
\midrule
 YC-1/4  & 0.5      & 0.7       &10  & 1   & 1     \\
 DIGI1   & 0.5      & 0.9       &10  & 0.5   & 2     \\
\midrule
 YC5      & 0.5      & 0.7       &10  & 1   & 1     \\
 DIGI5   & 0.7      & 0.7       &10  & 0.5   & 4     \\
 RR      & 0.5      & 0.9       &10  & 0.25& 4     \\
 NOWP    & 0.5      & 0.9       &10  & 1   & 1     \\
\bottomrule
\end{tabular}
\vspace{-3mm}
\end{table}

%% file: Figures/Fig6_train_size.tex
\begin{figure}
\centering

\begin{tabular}{c}
\includegraphics[width=0.40\textwidth]{Figures/Fig5_cropped_legend.pdf}
\end{tabular}\\

\begin{tabular}{cc}
\includegraphics[width=0.20\textwidth]{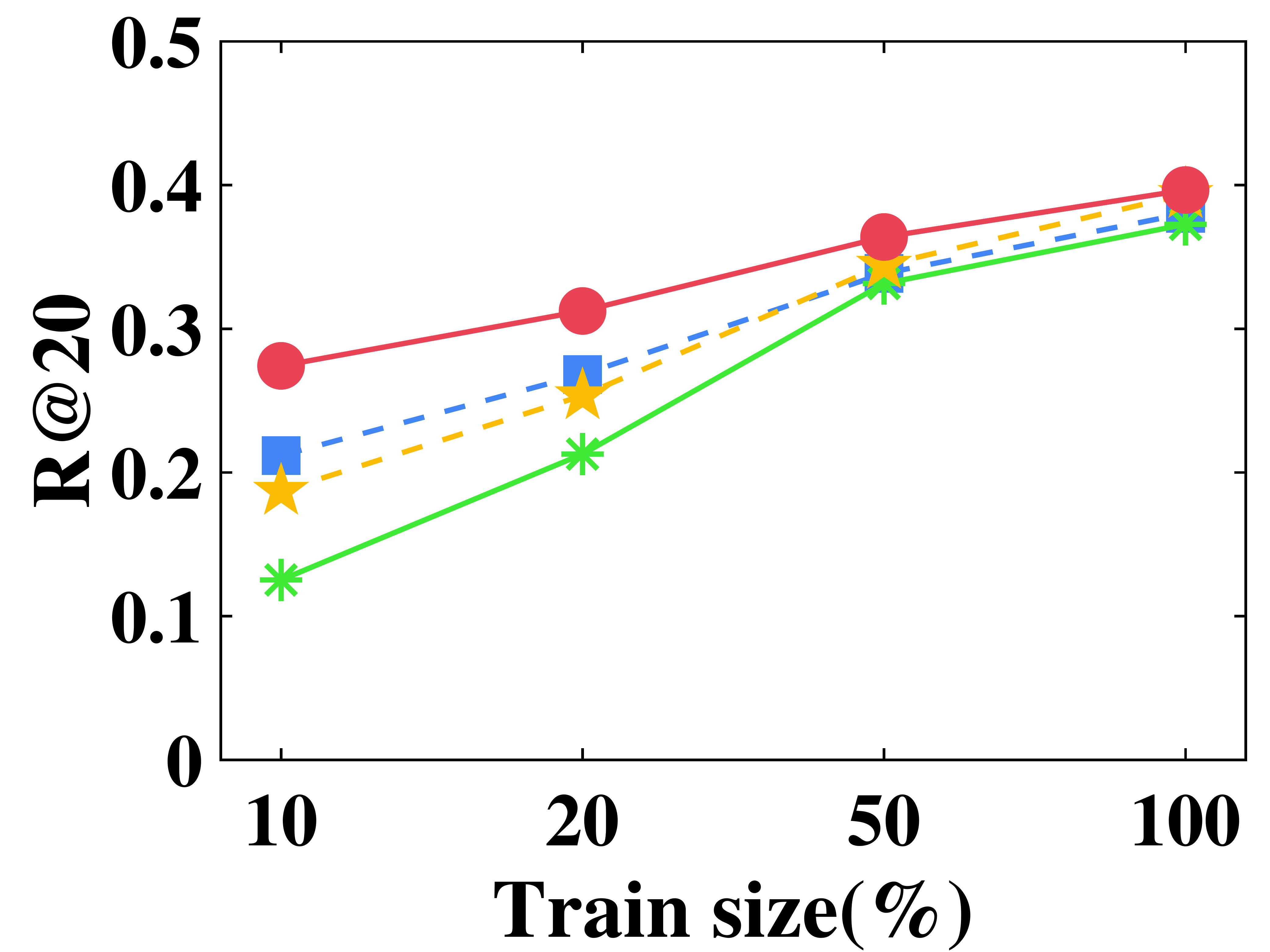} &
\includegraphics[width=0.20\textwidth]{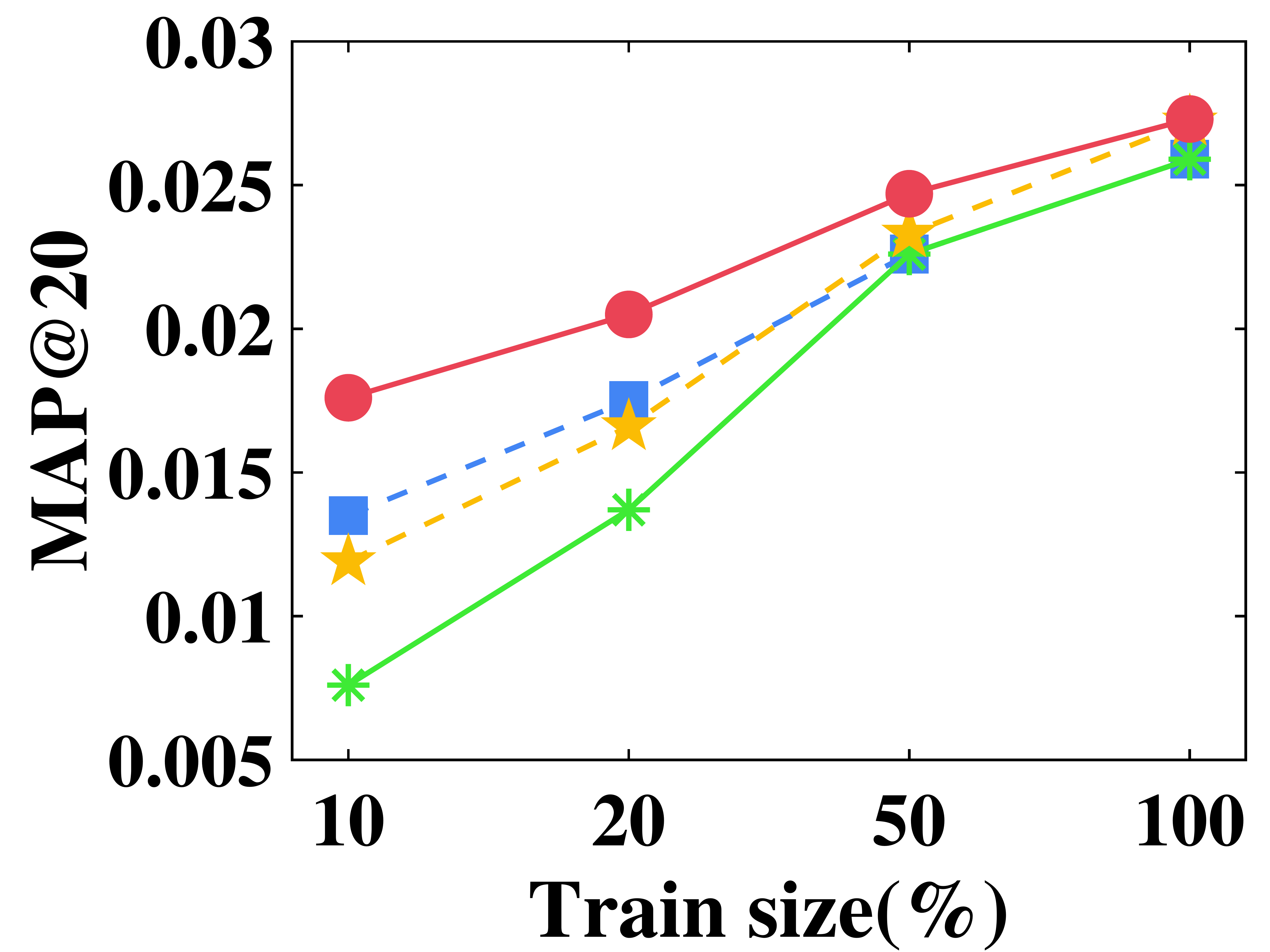}
\end{tabular}\\
(a) DIGI5  \\

\begin{tabular}{cc}
\includegraphics[width=0.20\textwidth]{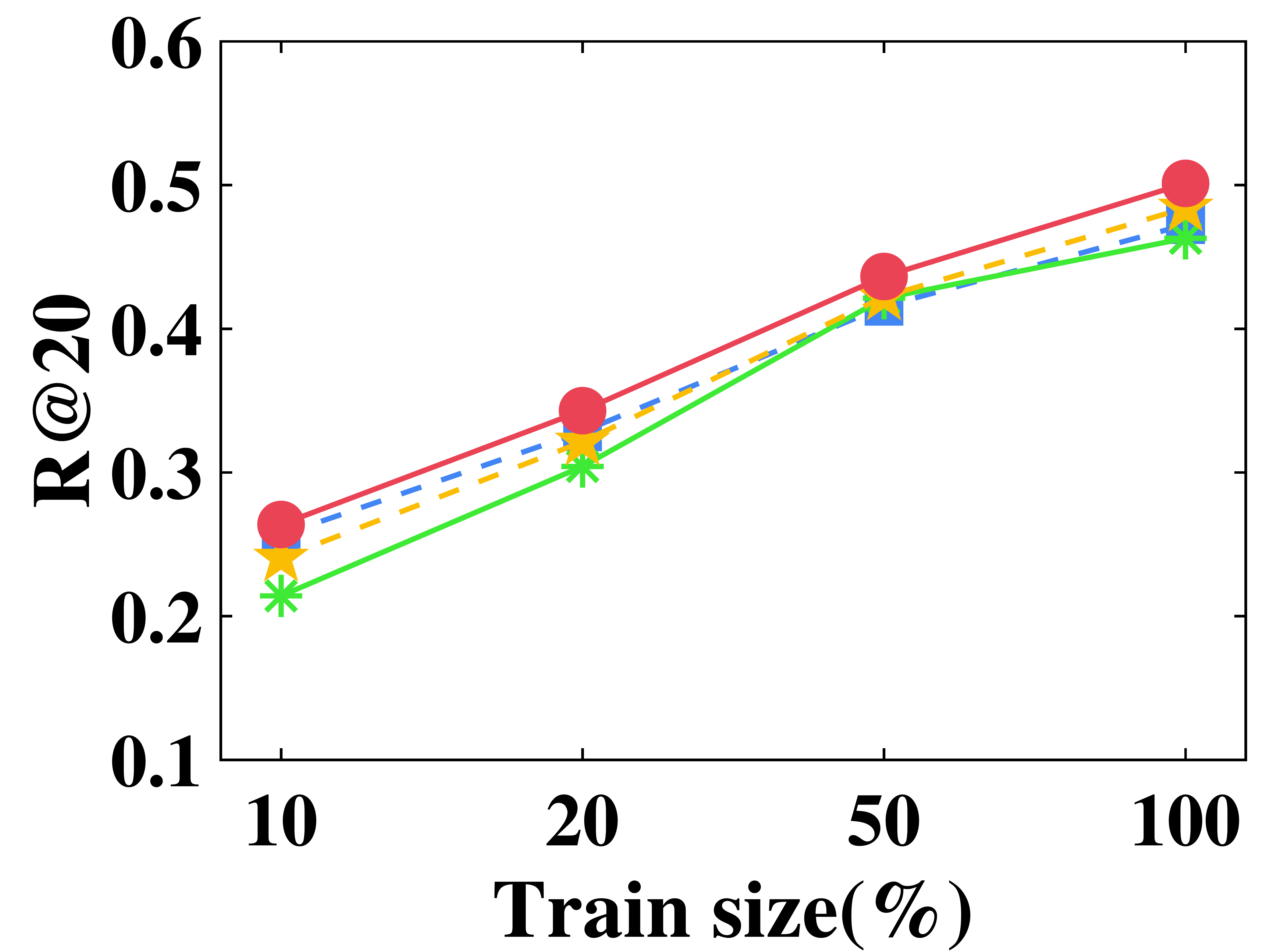} &
\includegraphics[width=0.20\textwidth]{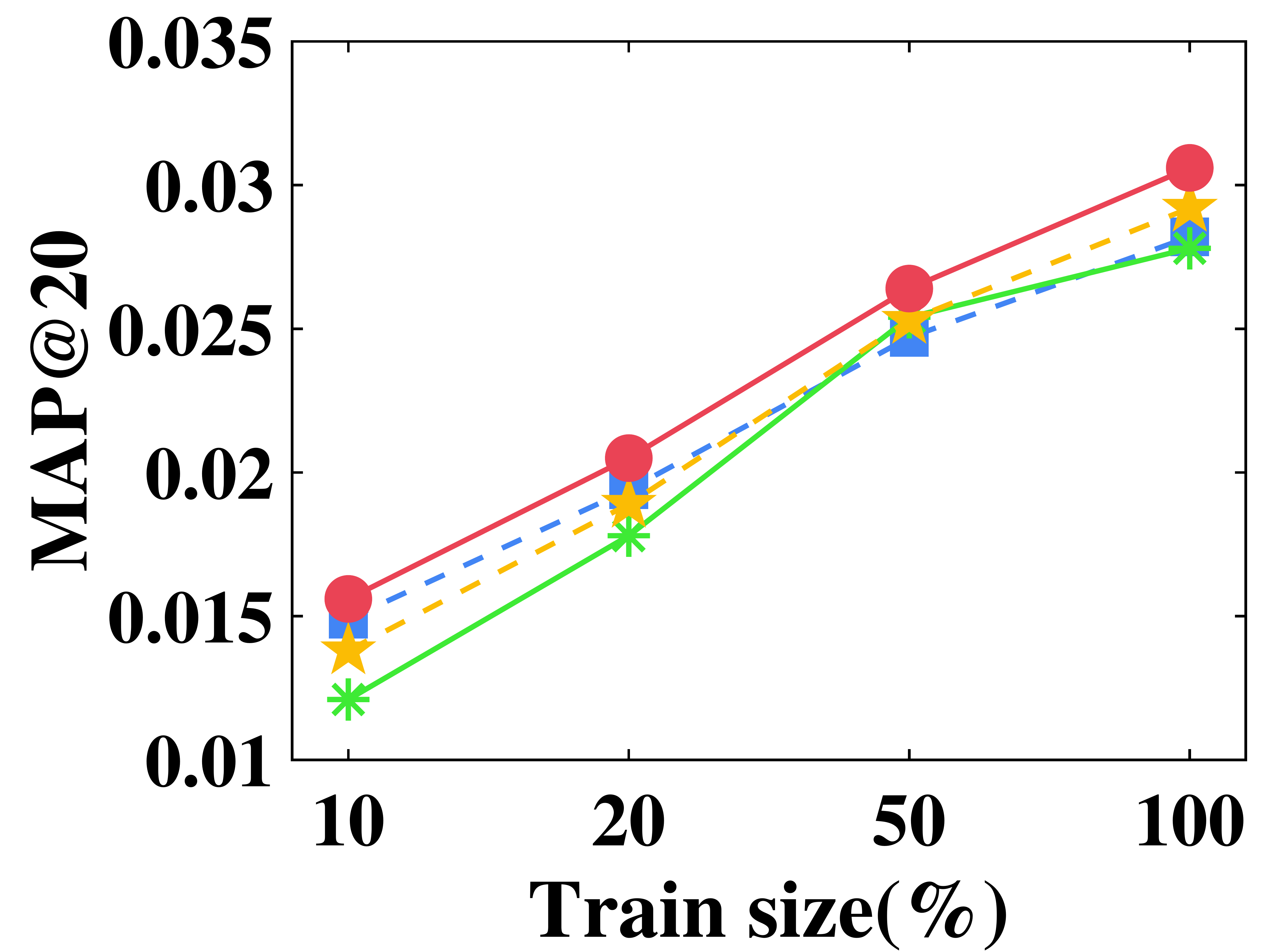}
\end{tabular} \\
(b) RR  \\

\vskip -0.1in
\caption{R@20 and MAP@20 of S-Walk and competing models over various size of training session matrices.}
\label{fig:train_size}
\vskip -0.15in
\end{figure}